\documentclass[12pt]{article}

\usepackage[top=4cm, bottom=4cm, left=3cm, right=2.5cm]{geometry}

\usepackage{setspace}
\usepackage{footmisc}
\makeatletter
\newcommand\iraggedright{%
\let\\\@centercr\@rightskip\@flushglue \rightskip\@rightskip
\leftskip\z@skip}
\makeatother
\iraggedright

\usepackage[utf8]{inputenc}
\usepackage[spanish, english]{babel}
\usepackage[LGR,T1]{fontenc}
\usepackage{float}
\date{}

\usepackage{physics}
\usepackage{natbib}
\usepackage{graphicx}
\usepackage{amssymb}
\usepackage{amsmath}
\usepackage{amsthm}
\usepackage{mathrsfs}
\usepackage{csquotes}
\usepackage[hidelinks]{hyperref}
\usepackage{enumerate}
\usepackage{calligra}
\usepackage{tikz}\usetikzlibrary{patterns, hobby}
\usepackage{epigraph}
\usepackage{verbatim}

\usepackage{titlesec}

\usepackage{array, tabularx}

\usepackage[T1]{fontenc}
\usepackage[osf,sc]{mathpazo}

\usepackage{xcolor}

\theoremstyle{definition}

\pgfdeclarepatternformonly{south west lines}{\pgfqpoint{-0pt}{-0pt}}{\pgfqpoint{3pt}{3pt}}{\pgfqpoint{3pt}{3pt}}{
        \pgfsetlinewidth{0.4pt}
        \pgfpathmoveto{\pgfqpoint{0pt}{0pt}}
        \pgfpathlineto{\pgfqpoint{3pt}{3pt}}
        \pgfpathmoveto{\pgfqpoint{2.8pt}{-.2pt}}
        \pgfpathlineto{\pgfqpoint{3.2pt}{.2pt}}
        \pgfpathmoveto{\pgfqpoint{-.2pt}{2.8pt}}
        \pgfpathlineto{\pgfqpoint{.2pt}{3.2pt}}
        \pgfusepath{stroke}}

\pgfdeclarepatternformonly{south east lines}{\pgfqpoint{-0pt}{-0pt}}{\pgfqpoint{3pt}{3pt}}{\pgfqpoint{3pt}{3pt}}{
    \pgfsetlinewidth{0.4pt}
    \pgfpathmoveto{\pgfqpoint{0pt}{3pt}}
    \pgfpathlineto{\pgfqpoint{3pt}{0pt}}
    \pgfpathmoveto{\pgfqpoint{.2pt}{-.2pt}}
    \pgfpathlineto{\pgfqpoint{-.2pt}{.2pt}}
    \pgfpathmoveto{\pgfqpoint{3.2pt}{2.8pt}}
    \pgfpathlineto{\pgfqpoint{2.8pt}{3.2pt}}
    \pgfusepath{stroke}}
    
\newcommand{\A}{{\mathcal A}}
\newcommand{\Hcal}{{\mathcal H}}
\newcommand{\B}{{\mathcal B}}
\newcommand{\Ocal}{{\mathcal O}}
\usepackage{authblk}

\usepackage{todonotes}

\newcommand{\counterfactual}{\ensuremath{%
		\mathrel{\Box\kern-1.5pt\raise1pt\hbox{$\mathord{\rightarrow}$}}}}

\usepackage[
]{geometry}

\DeclareMathOperator{\Id}{Id}

\title{The Causal Axioms of Algebraic Quantum Field Theory: A Diagnostic}

\author{Francisco Calderón\footnote{\href{mailto:fcalder\,@\,umich.edu}{\texttt{fcalder@umich.edu}}}}
\affil{\small Department of Philosophy, University of Michigan, Ann Arbor, MI, USA}

\begin{document}
\maketitle

Forthcoming in \textit{Studies in History and Philosophy of Science}

\begin{abstract}
Algebraic quantum field theory (AQFT) puts forward three ``causal axioms'' that aim to characterize the theory as one that implements relativistic causation: the spectrum condition, microcausality, and primitive causality. In this paper, I aim to show, in a minimally technical way, that none of them fully explains the notion of causation appropriate for AQFT because they only capture some of the desiderata for relativistic causation I state or because it is often unclear how each axiom implements its respective desideratum. After this diagnostic, I will show that a fourth condition, local primitive causality (LPC), fully characterizes relativistic causation in the sense of fulfilling all the relevant desiderata. However, it only encompasses the virtues of the other axioms because it is implied by them, as I will show from a construction by Haag and Schroer (\citeyear{Haag1962}). Since the conjunction of the three causal axioms implies LPC and other important results in QFT that LPC does not imply, and since LPC helps clarify some of the shortcomings of the three axioms, I advocate for a holistic interpretation of how the axioms characterize the causal structure of AQFT against the strategy in the literature to rivalize the axioms and privilege one among them.
\end{abstract}

% Research highlights
%\begin{highlights}
%\item 
%\item 
%\item 
%\end{highlights}

% Keywords
% Each keyword is separated by \sep

\maketitle

\vfill

\section{Introduction}

Philosophers have often praised algebraic quantum field theory's (AQFT's) mathematically precise, axiomatic framework. The theory's core is to assign to every open, bounded region of spacetime $\Ocal$ a set of observables $\A(\Ocal)$ describing the physics in said region. For this paper, every $\A(\Ocal)$ is a von Neumann algebra that, in particular cases, can be isomorphic to $\B(\Hcal)$, the bounded, linear operators on some Hilbert space $\Hcal$---a structure more familiar from ordinary quantum mechanics (QM). Although there has been much debate on which formulation of QFT we should use to address foundational, philosophical, and interpretive questions \citep{Wallace2011, fraser_how_2011, swanson_philosophers_2017}, arguably, the fruitfulness of choosing AQFT lies in the rich and robust framework that its axioms provide.
%However, as AQFT's advocates know, these axioms are very demanding from a mathematical and physical point of view, so we must be on the lookout for parts of the framework that are redundant (in some sense) or theoretically idle.

However, as AQFT's advocates know, these axioms are very demanding from a mathematical and physical point of view, so we must be attentive to the role each axiom plays in capturing the physical features we want a QFT to have. %However, as AQFT's advocates know, these axioms are very demanding from a mathematical and physical point of view. Since \textquote{We regard the postulates...as well as subsequent modifications and further structural assumptions as working hypotheses rather than as rigid axioms.} \citep[p. 58]{Haag1996}, we must pay attention to what role we want each axiom to play so that it captures the physical content we want a QFT to have.
It is then puzzling that AQFT lays down multiple axioms that prima facie look like they strive to do the same job---namely, characterizing the causal structure of a theory. %that we know is too demanding.
The spectrum condition (SC), microcausality (MC), or primitive causality (PC) are all deemed \textbf{causal axioms}. From this point of view, it's not surprising that the tendency in the literature on relativistic causation in QFT has been to adopt the following attitude:%The clearest evidence of their restrictiveness is the fact that the attempts to go beyond free fields have proven to be unsuccessful or require certain modifications \citep{Buchholz2020, Fredenhagen2015}. / Another troublesome instance of the worry of restrictiveness is the multiplicity of axioms of AQFT that have a causal gloss. Is it not overkill to lay down multiple axioms striving to do the same job---namely, characterizing the causal structure of a theory that we know is too demanding. / The tendency in the literature on relativistic causation in QFT has been to adopt the following attitude toward that question:

\begin{displayquote}
    \textbf{Atomistic conception of the causal axioms} (\textsc{Atomistic}): One of the axioms implements relativistic causation in (or characterizes the causal relativistic structure of) QFT singlehandedly or in the most exemplary way.
\end{displayquote}

The following passages are evidence of this attitude:

\begin{itemize}
    \item \textquote{[SC] is perhaps the most direct expression of the prohibition of spacelike processes} \citep[p. 303]{Butterfield2007}
    \item \textquote{The usual causality postulate which is adopted by practically all authors may be expressed in the following way...[MC]} \citep[p. 250]{Haag1962}.
    \textquote{the latter axiom [MC] represents the most important specific feature of relativistic systems} \citep[p. 14]{horuzhy_introduction_1990}. MC is still the single, most famous constraint among physicists, as most textbooks in conventional QFT can attest.
    \item \textquote{the most fundamental causality requirement in both classical relativistic field theory and relativistic AQFT is the requirement of no superluminal propagation. [SC] has sometimes been touted as a prohibition on superluminal processes, but we will argue that it is no such thing. The proper expression of the principle of no superluminal propagation in AQFT...is satisfied in models of AQFT that obey [PC].} \citep[p. 2]{EarmanValente2014}\footnote{I have renamed Earman and Valente's core causal condition to fit my terminology here. The reader should bear in mind that the literature on AQFT uses the same terms in different ways.}
\end{itemize}

The first main upshot of this paper is to reject \textsc{Atomistic}: none of these causal axioms \textit{fully} explains the notion of relativistic causation appropriate for AQFT, so we should not rivalize them and privilege one among them. As I will show, there are two kinds of shortcomings for defending \textsc{Atomistic}. First, the axioms only capture some (and not all) of the desiderata for relativistic causation that I will state in the next section. Second, it is often unclear how each axiom implements its respective desideratum.

The second main upshot is to advocate for the following attitude instead:

\begin{displayquote}
    \textbf{Holistic conception of the causal axioms} (\textsc{Holistic}): All of the axioms (and the connections between them) are required to implement relativistic causation in (or characterize the causal relativistic structure of) QFT.
\end{displayquote}

The cornerstone for \textsc{Holistic} will be a fourth condition, \textbf{local primitive causality} (LPC) (also called the \textbf{diamond property} in older literature). LPC will embolden \textsc{Holistic} in two ways. First, LPC \textit{does} fully characterize relativistic causation in the sense of fulfilling all the relevant desiderata. Second, it only encompasses the virtues of the other axioms because it is implied by them. More specifically, I prove that we can derive it from the rest of the axioms, which are known to be logically independent. While some readers might take LPC for granted, doing so would involuntarily assume Haag duality, a problematic assumption from the original derivation of LPC \citep{Haag1962}. Additionally, this would obfuscate the connections between the axioms. So, to reiterate, LPC grants us holism in two ways: by interpreting the axioms as mathematically rigorous but mostly physically intuitive and interpretatively rich constraints on a QFT, but also by taking them as members of an axiomatic \textit{system}.\footnote{As far as I know, there are no other explicit advocates of \textsc{Holistic}. Here are the two most similar claims I could find: \textquote{The principle of relativistic causality [MC] can be put into various mathematical forms and in terms of various objects: observables, states, elementary observables, corresponding to projection operators. And \textit{it would be hard to say that any particular formulation includes the total contents of this broad physical principle.}} (\citeyear[p. 20]{horuzhy_introduction_1990}, emphasis added). However, Horuzhy is discussing different ways of expressing the content of MC only. The other claim is the following passage from Rédei (endorsed by \citep[p. 17]{HSV2018}): \textquote{relativistic locality is \textit{not a single property} a physical theory can in principle have but an intricately interconnected web of features. Each of those individual features expresses some important aspect of relativistic locality, and a physical theory can in principle have some of these features but not others. A physical theory is in full compliance with relativistic locality if it possesses \textit{all} the features in this web} (\citeyear[p. 138]{redei_categorial_2014}). However, the crux of Rédei's holism is a unified framework from category theory. Not only can I make a similar claim with less sophisticated tools, but the holism brought about by LPC partially stems from fulfilling a set of physically motivated desiderata and not merely from belonging to a cohesive mathematical structure.} %\footnote{Here, I'm thinking about something like the distinction between ``soft'' and ``formal'' axioms from Rédei and Stöltzner (\citeyear{redei_soft_2006}). It's not an issue that the distinction is vague. Plus, some of the creators of AQFT blur this boundary: \textquote{We regard the postulates...as well as subsequent modifications and further structural assumptions as working hypotheses rather than as rigid axioms.} \citep[p. 58]{Haag1996}.}
As such, my worries driving this paper are very similar to those of the creators of AQFT: beyond their mathematical consistency, are these physically meaningful and well-motivated requirements? Are they interdependent, and if so, do they codify compatible or rival features of the theory?

The methodology I will adopt to address those questions is that of a diagnostic that scrutinizes what it means to be a \textit{causal} axiom. What this means is that to address a dispute between \textsc{Atomistic} and \textsc{Holistic}, we must go through each axiom aiming to take the \textquote{wide frame allowed by the general principles and, narrowing the...frame step by step,...look for the neuralgic spots.} \citep[p. 243]{haag_discussion_2010}. The main tools at our disposal are, to borrow the terminology from Peskin and Schroeder (\citeyear{Peskin1995}), the three ``themes'' of QFT: the concept of field, QM, and special relativity (SR). In other words, an axiom of AQFT should implement the notion of field (or, in other words, make clear what is \textbf{local} in the sense of situating phenomena in delimited patches of spacetime, as opposed to \textbf{global}) and have a clear input from QM and SR. Now, while Earman and Valente (\citeyear{EarmanValente2014}) have presented a similar overview and ended up with a form of \textsc{Atomistic}, another difference between their diagnostic and mine is a deliberate attempt to be minimally technical, partly because of how inaccessible the literature on AQFT is, and partly to make the philosophical stakes of these discussions more salient. But this is not to say that the diagnostic preceding my discussion of LPC is a mere pedagogical rehashing of the literature: aside from assuming less from the reader, I will introduce new difficulties for interpreting the axioms and surface different considerations for thinking about them. With this in mind, I've also relegated the gory details of the proof that the causal axioms imply LPC to an appendix and technical details that might distract readers from the argument to footnotes.

\section{Four desiderata for relativistic causation}

Before moving on with the diagnostic and seeing which axioms have which symptoms, I should clarify what ``relativistic causation'' means. At this point, I also depart from the literature on relativistic causation since the following desiderata are trying to track the folk notions of causation used by physicists. Contrast this approach with Butterfield's (\citeyear{Butterfield2007}), which focuses on the notions of causation that are widespread in the philosophy of science literature. Now, the physics literature has multiple facets of the notion of ``causation,'' so I will lay down multiple desiderata for relativistic causation.

First, as a form of ``action,'' relativistic causation attempts to capture a notion of (i) locality\footnote{``Locality'' is said in many ways. Specifically, it's sometimes used synonymously with ``localization,'' ``separability''/``independence,'' ``no superluminal signaling,'' or an obverse of Bell non-locality \citep[p. 3]{EarmanValente2014}. From this list, I will discuss the first three. However, I'm not using the term in the fourth sense, famously absent in quantum theories, especially in QFT.} and of (ii) a finite speed of propagation of events: \begin{displayquote}
For the relative independence of spatially distanced things (A and B), this idea is characteristic: external influences from A have no \textit{immediate} influence on B; this is known as the ``principle of proximal action [\textit{Nahewirkung}],'' which is only used consistently in field theory. The complete abolition of this principle would make impossible...the formulation of empirically testable laws in our familiar sense. \citep[p. 321-322, my translation]{Einstein1948}
\end{displayquote}
The worry about the \textit{immediacy} of the propagation of this kind of influence encapsulates (i) and (ii) into the preclusion of superluminal signaling, arguably the most famous notion of relativistic causation. Events are \textit{independent} in a relevant causal sense if we cannot connect them with a light signal. That is, we have merely specified what causal connections \textit{cannot} be. ``Relativistic causation'' also includes a notion of causal \textit{dependence}, not merely \textit{in}dependence \citep[p. 4]{EarmanValente2014}. Here is an account from another renowned physicist:
\begin{displayquote}
In physics, causal description...rests on the assumption that the knowledge of the state of a material system at a given time permits the prediction of its state at any subsequent time. \citep[p. 312]{Bohr1948}
\end{displayquote}
Therefore, (relativistic) causation should also capture an idea of (iii) a deterministic connection between events. Finally, I will add (iv) respecting the metric structure of the underlying spacetime since Geroch (\citeyear{Geroch2011}) has shown that SR is a viable theory even with superluminal signaling and since SR secures very few of the multiple locality conditions of AQFT (and the ones implementing SR are independent of the rest) \citep[p. 314]{Ruetsche2021}. Additionally, some axioms are restricted to Minkowski spacetime while others only require that the spacetime is globally hyperbolic, so different metric structures avail different resources to interpret the content of the axioms.

\section{Symptomatology}

\subsection{The Spectrum Condition}

The first axiom we will consider is the spectrum condition (SC):
\begin{displayquote}
\textbf{SC}: (a) The joint spectrum of the infinitesimal generators of translations in Minkowski spacetime is confined to the forward lightcone. (b) There is a unique (up to phase factors) \textbf{vacuum state}, which is translationally invariant and has zero energy-momentum.
\end{displayquote}
In other words, we will only consider states with positive energy-momentum (plus the vacuum). SC is usually supplemented with a local ``counterpart,'' the \textbf{axiom of covariance}, which dictates how the algebra of a given region transforms under the unitary representations of Lorentz transformations. Since SC and the axiom of covariance cover the demands of the Poincaré group of SR, including both Lorentz transformations and spacetime translations, the axiom of covariance is usually taken as one of the causal axioms of the theory. Now, from this perspective of symmetries, they are both straightforward applications of Wigner's theorem: to obtain the quantum analog of the symmetry group of a classical theory (in this case, the Poincaré group), take the group's unitary representations. Thanks to Stone's theorem, those unitary representations can be further expressed in terms of their infinitesimal generators, which, for the translation subgroup, are the key ingredient for SC. So, the way in which SC implements QM and SR only codifies the geometry of Minkowski spacetime. That covers the fourth desideratum: SC respects the structure of (Minkowski) spacetime. What about the other desiderata? And is it not a shortcoming of SC as a causal axiom that we surely want to be able to claim that there is some form of ``relativistic causation'' in AQFT in curved spacetimes?

Let's start with the first question. SC is fairly silent about determinism, and although it's not local, the axiom of covariance is. As for forbidding superluminal signaling, its role (if it has one) needs to be clarified. SC does not hold for every QFT since the energy (density) of a \textit{quantum} field need not be positive \citep{Epstein1965}. On top of that, the energy-momentum operators do not necessarily coincide with the energy-momentum (tensor) of the quantum field in question \citep{EarmanValente2014}. Now, SC doesn't hold for every QFT, but quantum fields obtained by quantizing classical relativistic fields exhibit no superluminal signaling \citep[pp. 17-18]{EarmanValente2014}. Maybe there's something about field theories in general precluding superluminal signaling? It doesn't seem that way since the classical analogs of SC are not sufficient conditions to prohibit superluminal signaling \citep[pp. 103-105]{Earman2014}. As for the second question, SC can be stated at every point of a curved spacetime \citep[p. 338]{Haag1996}, but  (non-unique) generalizations of restrictions on the energy-momentum of the fields \citep{curiel_primer_2017} are extensions from the classical analogs of SC I just mentioned are not sufficient to prohibit superluminal signaling.

If SC (a) does not fulfill the goal we would want for \textit{a single} causal axiom to have or to have in the most exemplary way, what about SC (b)? SC (b) is sometimes taken not as a characterization of relativistic causation in AQFT but, ironically, as one of the ingredients needed to question AQFT's status as a relativistic theory. One of the consequences of SC (when supplemented with a condition of ``additivity'' saying that observables can be expressed in terms of observables of arbitrarily small regions) is the \textbf{Reeh-Schlieder theorem}. The theorem claims that if we act on the vacuum defined by SC (b) with an element of $\A(\Ocal)$ for some region $\Ocal$, we can generate any other state. This result is unsettling because it seems that we can measure the energy of the vacuum in a lab and yet approximate any other quantum state in any other region of spacetime, even states that \textit{do} have energy! There are interpretive strategies and more sophisticated mathematical tools to get out of this problem \citep{Halvorson2001, wallace_defence_2006, valente_does_2014}. Still, they do not change that SC (b) does not implement the prohibition of superluminal signaling, even if we manage to show that it does not impede it.

\subsection{Microcausality}

The second axiom we will consider is the microcausality (MC), Einstein causality, or local commutativity condition:
\begin{displayquote}
\textbf{MC}: $\A(\Ocal_1)$ commutes with $\A(\Ocal_2)$ if $\Ocal_1$ and $\Ocal_2$ are space-like separated.
\end{displayquote}
Let us unpack this condition: since no physical process can occur along a spacelike trajectory, no measurement in $\Ocal_1$ can disturb the outcomes of a measurement carried out in $\Ocal_2$, and vice-versa. The notion of spacelike separation is tracking the metric structure of spacetime (and can be extended to locally flat tangent spaces in curved spacetimes \citep[p. 338]{Haag1996}), and this condition is clearly local. Now, MC is often rephrased to say that there would be statistical correlations if we could connect $\Ocal_1$ and $\Ocal_2$ with a superluminal signal. However we want to read MC, it's an attempt to transcribe to the language of operator algebras the passage I cited from Einstein in the introduction. For that reason, MC is considered a condition of \textbf{independence} or \textbf{separability} between quantum systems in regions $\Ocal_1$ and $\Ocal_2$.

Prima facie, MC is a straightforward implementation of relativistic causation in the sense of no superluminal signaling. That is, MC attempts to capture the first two desiderata of relativistic causation that I had laid out. Aside from only capturing some of the desiderata for relativistic causation in AQFT, this section will show that MC's attempt to prohibit superluminal signaling comes with a \textquote{morass of recalcitrant interpretational issues.} \citep[p. 16]{EarmanValente2014}.

Since the way in which this axiom implements its causal desiderata comes off badly from my diagnostic, I want to stress its importance for QFT in general before moving on. First, MC can be easily corroborated for the most common field theories, or it is a crucial assumption in their construction. Consider a Klein-Gordon field, the simplest relativistic free field. Then, the commutator of quantized fields operators at spacetime points $x$ and $y$ vanishes for equal times $x^0$ and $y^0$ (for any spatial separation $\Vec{x}-\Vec{y}$) and spacelike values of $x-y$. Second, MC can be adapted and modified to include fields with spin or charge, like the Dirac field, via the spin-statistics theorem. It is even a sufficient condition to define a Lorentz-covariant scattering matrix since it is connected to important analyticity properties of the fields \citep{duncan_dynamics_2012, Weinberg1995}. It is, then, no surprise that physicists follow Einstein's suit in putting MC at the heart of relativistic causation in QFT.

However, MC is not as straightforward as it seems. For it to be a QM-informed axiom, we must now bite the bullet of not knowing what ``measurement'' means in a quantum setting \citep[p. 11]{EarmanValente2014}. Whatever ``measurement'' means, we need observables and states to obtain expectation values. One way to obtain the theory's observables is through a quantization procedure of a classical field theory.\footnote{Since the fields are operator-valued distributions, we also need a sufficiently smooth test function to smear them before using them in calculations.} Then, we need a state $\ket{\Psi}$ such that $AB\ket{\Psi}=BA\ket{\Psi}$, where $A\in\A(\Ocal_1)$ and $B\in\A(\Ocal_2)$ for $\Ocal_1$ and $\Ocal_2$ are spacelike separated. Since $\Ocal_1$ and $\Ocal_2$ are subregions of the background spacetime $\mathcal{M}$, $\A(\Ocal_1)$ and $\A(\Ocal_2)$ are subalgebras of $\A(\mathcal{M})$. Given this property of \textbf{isotony} and since $\Ocal_1$ and $\Ocal_2$ are spacelike separated, $\A(\mathcal{M})$ should admit a state. However, it would be absurd that local operations like $A\ket{\Psi}$ or $B\ket{\Psi}$ would alter a physical state for all space and time \citep[p. 110]{Ruetsche2011}.

Despite these problems, MC still seems to be a reasonable axiom for a field theory in implementing some ``separability'' of physical subsystems, which, additionally, has testable consequences in terms of the measurements we can perform in each subsystem. A reason to probe this further is that, following Einstein's quotation, the separability of physical subsystems and the possibility of communicating them are closely connected, so we need a better grasp of that connection before diagnosing MC as a causal axiom. I will address those topics separately.

%\vspace{-1em}

\subsubsection{Separability/Independence conditions}

This subsection aims to nail down some interpretive difficulties of diagnosing MC as an axiom in AQFT. Since MC strives to make the separability of quantum field theoretical systems compatible with the prohibition of superluminal signaling between them, this subsection sets the stage for my discussion of causation in the following subsection.

Consider $\A(\Ocal_1)\vee\A(\Ocal_2)$, the smallest algebra generated by $\A(\Ocal_1)$ and $\A(\Ocal_2)$, where $\Ocal_1$ and $\Ocal_2$ are spacelike separated. A state $\omega$ in AQFT is a linear, positive-definite functional over the operator algebra. For our purposes, $\omega(A)$ is the expectation value of the observable $A$ belonging to some algebra $\A$. Now, consider a pair of states $\omega_j$ acting on $\A(\Ocal_j)$ for $j=1,2$. Then, we should be able to find a state $\omega$ acting on $\A(\Ocal_1)\vee\A(\Ocal_2)$ such that its restriction to $\A(\Ocal_j)$ is just $\omega_j$. To give an illustration more familiar to readers coming from QM, if the algebras are matrix algebras, the composition $\vee$ can be equivalent to a tensor product, and the expectation value $\omega(A)$ can be calculated using the trace prescription for a density matrix $\rho_\omega$. In that case, the ``restriction'' from the global state to the local states would be taking the partial trace over either one of the subsystems.

The most famous form\footnote{Earman and Valente (\citeyear[§4-5]{EarmanValente2014}) survey other ways to formulate independence conditions.} of independence in the spirit of my last paragraph is \textbf{statistical independence}, which is merely taking the expected values of $\omega_1$ and $\omega_2$ as probabilistically independent, that is, $\omega (C) = \omega_1(C)\omega_2(C)$ for some observable $C\in\A(\Ocal_1)\vee\A(\Ocal_2)$. There is no resulting statistical mixture of the individual states because they are localized in causally disjoint regions. This form of independence looks like a straightforward rendition of what we would want the compatibility of measurements carried out in spacelike separated regions to be since we are breaking potential correlations between the measurements in both regions. However, for statistical independence to hold, we need an additional assumption about the nature of the algebras called the \textbf{split property}, which in turn depends on the fields' energy densities having certain properties \citep{Fewster2016} different from SC that, though reasonable, are not always fulfilled. That is, it can be shown that statistical independence can be derived from MC, but it needs additional assumptions. On their own, MC and statistical independence are logically independent \citep[p. 11]{EarmanValente2014}. Moreover, more general forms of independence that avoid superluminal signaling do not imply MC \citep[pp. 753-755]{Halvorson2007}.

Still, MC does seem to implement some ``separability'' of physical subsystems, even if it is not because of their independent statistical predictions. One suggestion is to see MC more as a form of mereology of the structure of the algebras than a claim about locality \citep[pp. 112-113]{Ruetsche2011}; the operation $\vee$ in $\A(\Ocal_1)\vee\A(\Ocal_2)$ should incorporate and respect MC. However, consider the composition of physical subsystems of standard QM: take two Hilbert spaces $\Hcal_1$ and $\Hcal_2$, and construct their tensor product, $\Hcal_1\otimes\Hcal_2$. Then the algebras $\B(\Hcal_1)\otimes\Id_2$ and $\Id_1\otimes\B(\Hcal_2)$ commute, and yet QM is not a relativistic theory. This example shows that the composition of bigger subsystems from commuting algebras is not a feature of relativistic theories \citep[p. 752]{Halvorson2007}. So even if MC seems reasonable, the way in which it combines QM and SR brings to light multiple difficulties.

%\vspace{-1em}

\subsubsection{MC and causation}

Aside from statistical independence, the \textbf{no-signaling theorem} is a result closely connected to MC, but that shifts our focus from separability to relativistic causation. In this subsection, I will finally show that the problems of separability and measurements in the two previous ones disrupt the possibility that MC fully characterizes relativistic causation in AQFT.

Of the many formulations of the no-signaling theorem, I will follow \citep[§4.2]{EarmanValente2014} to avoid further technicalities. Consider an observable $A\in\A(\Ocal_1)$ with a (countable) spectral decomposition $A=\sum_i a_i E^A_i$, where the $E_i$s are projectors and the $a_i$s $A$'s eigenvalues. We can then define a map $T^A(\cdot)=\sum_i E^A_i \cdot E^A_i$ over $\A(\Ocal_1)\vee\A(\Ocal_2)$ that is explicitly related to an $A$-measurement. Then, for a state $\omega$ and an observable $B\in\A(\Ocal_2)$, we have:
\begin{equation*}
    %\begin{aligned}
    \omega(T^A(B)) = \omega(B) %&= \omega(\sum_i E^A_i B E^A_i) \overbrace{=}^{\text{Linearity of $\omega$}} \sum_i\omega(E^A_i B E^A_i) \\ &\overbrace{=}^{\text{MC: $[A,B]=0$, so $[E^A_i,B]=0$}} \sum_i\omega(B E^A_i E^A_i) \\ &\overbrace{=}^{\text{$(E^A_i)^2=E^A_i$}} \sum_i\omega(B E^A_i) \overbrace{=}^{\text{Linearity of $\omega$}} \omega(B \sum_i E^A_i) \overbrace{=}^{\text{$\sum_i E^A_i=\Id$}} \omega(B)
    %\end{aligned}
\end{equation*}
This means that states acting on the algebras of spacelike separated regions are unaffected by measurements in the other region. The moral is clear: MC allows the statistical predictions of one of the subsystems to be preserved even after performing measurements in the other.

However, a few complications stop us from proclaiming victory over superluminal signaling using MC. First, the preservation of the statistics of the outcomes of a measurement may not require commutativity after all \citep{RedeiValente2010}, and the no-signaling theorem only works for a restricted class of operations \citep[p. 111]{Ruetsche2011}. Second, even if we have no-signaling theorems, the statistical independence of subsystems can get bypassed by measurement protocols like the one from figure \ref{sorkin}.
\begin{figure}[h]
\centering
%0.3
\begin{tikzpicture}[scale=0.45]
\draw[->] (-7,-3) -- (-7,3);
\draw (-7, 0) node[left] {Time};
\draw (-6,-6) to[curve through={(-5,-4.5) .. (-4,-5) .. (-3,-5.5) .. (-3,-6)}] (-6,-6);
\draw (9,5) to[curve through={(8,3.5) .. (7,4) .. (6,4.5) .. (6,5)}] (9,5);
\draw (-3,0) to[curve through={(-2,1) .. (-1,1.5) .. (1,2) .. (4,1.5) .. (6,0) .. (4, -2) .. (2, -1.5) .. (0, -2) .. (-1, -2.5)}] (-3,0);
\draw[blue] (-3,-6) -- (-1.3,-2.6);
\draw[blue, dashed] (-1.3,-2.6) -- (1,2);
\draw[blue] (1,2) -- (3,6);
\draw (-4.5,-6) node {$\Ocal_1$};
\draw (1.5,0) node {$\Ocal_2$};
\draw[blue] (0.5,-6) -- (2.7,-1.6);
\draw[blue, dashed] (2.7,-1.6) -- (4.25,1.5);
\draw[blue] (4.25,1.5) -- (6,5);
\draw (7.5,5) node {$\Ocal_3$};
\end{tikzpicture}
\caption{$\Ocal_1$ and $\Ocal_3$ are spacelike separated.}
\label{sorkin}
\end{figure}
For this setup, successive observations of the same state in $\Ocal_1$, $\Ocal_2$, and $\Ocal_3$ exhibit correlations between the measurement carried out in $\Ocal_1$ and that in $\Ocal_3$ \citep{Sorkin1993}. This result seems to imply that there is some superluminal signaling in QFT. Although Sorkin's paradox just turned 30, it has been significantly overlooked by most people working on measurements in QFT until a couple of years.\footnote{For a recent exception, see \citep{papageorgiou_eliminating_2023}.} Some papers claim to have found a solution \citep{Bostelmann2021}, but whether it is satisfactory or not, it is more contrived than assuming MC.

In conclusion, although MC seems to shed some light on relativistic causation in AQFT, multiple difficulties impede us from claiming we have achieved a complete, clear characterization. First, as an axiom stated in terms of quantum measurements, we fall into the difficulties of understanding what ``measurement'' means and, especially, what it means for spacelike measurements to be compatible. We tried to implement MC as a form of separability for this purpose. Still, statistical independence needs more than MC to work, and some other forms of independence are logically independent of MC or satisfied in non-relativistic cases. Even having failed to characterize what form of separability MC implies, the no-signaling theorem made us hopeful that it would still prohibit superluminal signaling. However, the theorem only works for a particular case of measurements, it can be proved without MC, and it can be bypassed in setups like Sorkin's. The takeaway is that we still need a good characterization of relativistic causation in AQFT. Every time MC seemed to lead us in the right direction, interpretive difficulties or other pitfalls stopped us from clearly seeing the role of MC in prohibiting superluminal signaling.

\subsection{Primitive Causality}

From my initial characterization of relativistic causation, SC and MC have aimed to implement the role of locality, subluminality, and the metric structure of (Minkowski) spacetime. However, I have not discussed determinism. Primitive causality (PC), or the time slice axiom, will fill this gap. First, we need some definitions. Consider two spacetime regions $\Ocal_1$ and $\Ocal_2$. $\Ocal_2$ \textbf{depends causally} on $\Ocal_1$ if every light ray in the backward (or forward) light cone originating from any point in $\Ocal_2$ intersects $\Ocal_1$.
\begin{figure}[h]
\centering
%0.3
\begin{tikzpicture}[scale=0.36]
\draw[->] (-9.5,0) -- (-9.5,7);
\draw (-9.5, 3.5) node[left] {Time};
\draw (-7,0) to[curve through={(-1,-1.5) .. (1,-1) .. (3,-2) .. (6,1.5) .. (3,2.5) .. (1,3) .. (-1,2.5) .. (-3,4)}] (-7,0);
\draw (-5,7) to[curve through={(-4,5) .. (-2,5) .. (1,6) .. (3,5) .. (4,7) .. (3,9) .. (-1,8) .. (-4,8)}] (-5,7);
\draw[blue] (-4,11) -- (-5,7) -- (-7,-1);
\draw[blue] (-6,11) -- (-5,7) -- (-3,-1);
\draw[blue] (2,-1) -- (4,7) -- (5,11);
\draw[blue] (3,11) -- (4,7) -- (6,-1);
\draw (8,0) node {$\Ocal_1$};
\draw (6,7) node {$\Ocal_2$};
\end{tikzpicture}
\caption{$\Ocal_2$ depends causally on $\Ocal_1$.}
\label{dependence}
\end{figure}
Then we have:
\begin{displayquote}
\textbf{PC}: If $\Ocal_2$ depends causally on $\Ocal_1$, $\A(\Ocal_2) \subseteq \A(\Ocal_1)$ (cf. figure \ref{dependence}).
\end{displayquote} Notice that if $\Ocal_2\subset\Ocal_1$, $\Ocal_2$ depends causally on $\Ocal_1$, so the property of isotony can be derived from PC. Additionally, since $\mathcal{M}$ depends causally on any time slice and any time slice is isotonically included in $\mathcal{M}$, PC is usually presented in the literature in an alternative way:
%Although I will not use the term ``local primitive causality'' to refer to PC, it is often called that way to contrast it with the following condition, which is how Haag et al. usually present primitive causality (\citeyear{Haag1962, Haag1996}):
\begin{displayquote}
Consider a \textbf{time slice} in Minkowski spacetime.\footnote{More generally, the spacetime manifold should be globally hyperbolic, and the time slice is some Cauchy surface.} That is, a region infinitely extended in space but restricted to a time interval of size $\tau$: $\Ocal_{\tau}:= \{x\in\mathcal{M}: \abs{x^0 - t} < \tau \}$. Then $\A(\Ocal_{\tau})=\A(\mathcal{M}) \ \forall\tau$.
\end{displayquote}
%$\mathcal{M}$ depends causally on any time slice, and any time slice is isotonically included in $\mathcal{M}$, so both versions end up being equivalent. However,
This version of the axiom makes determinism à la Bohr even more salient since the standard interpretation of PC is that there should be a ``dynamical law'' that allows us to determine the values of the fields at any given time with the values of the fields at a time slice \citep[p. 48]{Haag1996}. This claim can be re-stated in the following way: the time slice provides a region of evaluation of an initial value problem, and the ``dynamical law'' is deterministic \citep{Bogolubov1990}. Clearly, the exact nature of the law depends on the dynamics of the fields \citep[p. 21]{EarmanValente2014}. Since deterministic laws have causal-like behavior, PC is the core of Earman and Valente's \textsc{Atomistic} view. This section will argue that PC's determinism is insufficient to characterize relativistic causation in AQFT.

As mentioned, PC is local, and its time slice version is global. Aside from this point, I will treat both versions of PC interchangeably. Dimock proved the relevance of PC for QFT since it holds for the Klein-Gordon field (\citeyear{Dimock1980}) and the Dirac field (\citeyear{Dimock1982}). However, as an initial value problem, it is trivially fulfilled. The Lagrangians of field theories are formulated in terms of the fields and their derivatives at a time slice, and if their derivatives are of a sufficiently higher order, we can recover the field values in all spacetime \citep[pp. 330-331]{Bogolubov1990}. As such, PC is an entirely reasonable assumption for AQFT to make, but only because this form of ``determinism'' is a truism for QFT.\footnote{Within AQFT, there are technical difficulties that complicate claiming that we can restrict the global dynamics to those of a time slice. So, if it is a truism in Lagrangian QFT, it is hardly obvious that it holds in AQFT. Thanks to Noel Swanson for this remark.} PC is not, then, a feature of relativistic causation \textit{in AQFT} but in field theories in general.

Now, the QFTs considered by Dimock should satisfy PC since their equations of motion are hyperbolic partial differential equations (PDEs), which are characterized by having unique solutions within their domain of dependence \citep{Geroch2011} and by having a finite propagation speed \citep{Bar2007}. The notion of \textbf{domain of dependence} has a precise definition within the realm of PDEs, but, in our case, it is merely the locus of causally dependent points to some region $\Ocal$, denoted by $D(\Ocal)$. Another example of a hyperbolic PDE is the wave equation describing every undulatory phenomenon. From this point of view, PC is not a requirement from SR since it appears in non-relativistic phenomena, nor QM since it appears in classical ones. It is also not a sufficient characterization of relativistic causation since wave equations like the Klein-Gordon one would only be relativistic because their finite speed of propagation is the speed of light, but that does not rule out any higher speeds! These are the same reasons why PC is not a good causal axiom: the claim that the dynamics of the field are deterministic does not mean that the theory is causal in the sense that the dynamics should \textit{also} respect the light cone structure of the underlying spacetime \citep[p. 19]{HSV2018}.

Additionally, as Earman (\citeyear{Earman2014}) has claimed, determinism should apply not only to the time evolution of the field observables but also to the states. If states $\omega_1$ and $\omega_2$ give the same predictions with the observables in $\A(\Ocal)$, they should also provide the same ones with the observables in $\A(D(\Ocal))$. Although I agree with Earman's suggestion, he bases it on taking $\A(\Ocal)=\A(D(\Ocal))$ as a result of Haag and Schroer (\citeyear{Haag1962}). However, this equation is not valid in general. We only obtain $\A(\Ocal)\subseteq\A(D(\Ocal))$ thanks to PC, and that is different from what Haag and Schroer prove, which is a consequence of the local primitive causality that I will discuss in the next section.

\subsection{Taking stock: against \textsc{Atomistic}}

In the past subsections, we have seen the shortcomings and interpretive problems with the three causal axioms. Some of their drawbacks stem from not implementing QM, SR, and locality in a way amenable to codifying the desiderata of relativistic causation I laid out or from implementing them in ways that imported interpretive difficulties from each of those themes. And yet, someone who endorses \textsc{Atomistic} might be undeterred: if none of these causal axioms do the job, let us look for a condition that puts all those advantages in one place! In the next section, I will argue that local primitive causality (LPC) does that for us, but I will show that it does so because it is implied by the conjunction of SC, MC, and PC, thus going against \textsc{Atomistic}.

\section{LPC as a defense of \textsc{Holistic}}

To state LPC, we first need to define the \textbf{causal complement} of a region $\Ocal$, $\Ocal'$, as the set of points outside $D(\Ocal)$. Then we have:

\begin{displayquote}
\textbf{LPC}: For a region $\Ocal\subset\mathcal{M}$,\footnote{\citep[p. 19]{HSV2018} give a more general definition that I want to give here. See also their footnote 2.}
$\A(\Ocal)=\A(\Ocal'')$.
\end{displayquote}

I will use LPC to defend \textsc{Holistic} in two ways. First, contrary to the causal axioms, I will show that it satisfies all the desiderata for relativistic causation. In summary, LPC is local since its dependence on spacetime regions is explicit; it is a crucial ingredient in showing that concrete experimental setups exhibit no superluminal signaling \citep{Buchholz1994, Yngvason2005}; it is the key assumption in theories of local observables that attempt to ensure that the determinism of PC is compatible with the temporal evolution of field values in the light cone of their regions \citep{HSV2018}; and it emphasizes the metric structure of $\mathcal{M}$ underlying the definition of ``causal complement.'' I will only elaborate on how LPC rules out superluminal signaling and improves upon PC's attempt at capturing determinism.

However, LPC's role of unifying the desiderata does not warrant promoting it to the ring to rule all the causal conditions of AQFT. The second defense of \textsc{Holistic} will come as proof that the three causal axioms imply LPC, thus showing that LPC crystallizes how all the axioms hang together. Although the first version of this proof was first suggested by Haag and Schroer (\citeyear{Haag1962}) and I use their diamond construction, I fill in the missing steps and amend it in the appendix by not using Haag duality, which people involuntarily commit to when using LPC, but it is not satisfied by some of our best field theories and only works in particular spacetime regions \citep[pp. 145-147]{Haag1996}, thus restricting greatly the kind of physical models we can consider.

\subsection{LPC and determinism}

The time slice version of PC implements determinism too crudely: if we want to know whether some event involving quantum fields causes another one, we specify the values of the fields (and probably some of its derivatives) at \textit{a whole time slice}, and the pay-off is \textit{every possible operation on spacetime}. As argued by Hofer-Szabó and Vecsernyés (\citeyear{HSV2018}), the axiom of covariance allows us to define a time-evolution for the fields (similarly to how we implement time translations unitarily in QM) $\alpha_{\tau',\tau}$ taking the fields from time slice $\tau$ to $\tau'$. And yet, PC can't guarantee that $\alpha_{\tau',\tau} (\A(\Ocal_\tau))\subseteq\A(\Ocal_{\tau'}\cap\mathcal{D}(\Ocal_\tau))$! That merely symbolizes what I mentioned in the PC section: the dynamical law does not confine the time evolution of the fields to their light cones.

LPC grants us a more fine-grained notion of determinism, especially since it's formulated for finite regions. Since we assume that the underlying spacetime is globally hyperbolic, there will be a Cauchy surface transversing every open, bounded region $\Ocal$. What LPC tells us, then, is that specifying the values of the fields, not in a whole Cauchy surface but in the chunk intersecting $\Ocal$ suffices to know that operations in $\A(\Ocal)$ can influence operations in $\A(\Ocal'')$. Figure \ref{diamond} below will help illuminate this point: since the causal future of $\mathcal{C}$ intersects the causal future of points in $\mathcal{C}'$, claiming that $\A(\mathcal{C})=\A(\mathcal{C}'')$ tells us how much can we enlarge the operations we do in $\mathcal{C}$ while still being insulated from possible influences from operations in $\mathcal{C}'$. So, fulfilling Hofer-Szabó and Vecsernyés's aspirations of a condition that is \textit{both} deterministic and restricted to the light cones of finite regions (as opposed to infinite time slices), LPC is the cornerstone of a local, causal dynamical evolution. In the limit where $\mathcal{C}$ extends infinitely spatially, a full Cauchy surface transverses it, and the diamond $\mathcal{C}''$ becomes the full spacetime, as expected from PC! 

\subsection{LPC and superluminal signaling: The Fermi's Two-Atom System}

In this subsection, I will introduce (but skip most of the details) an experimental setup devised by Fermi (\citeyear{fermi_quantum_1932}) to check whether the quantum theory of radiation that he was studying was compatible with SR and translated in \citep{Yngvason2005} to an AQFT language. Consider two atoms, $a$ and $b$, separated by a distance $R$. At time $t=0$, $a$ is in its ground state, and $b$ is in an excited state. When $b$ emits radiation, there is a non-zero probability that $a$ is excited for $t>0$. Since the effect of $b$'s decay cannot be superluminal, $a$ should be in its ground state for $t<R/c$, where $c$ is the speed of light. (See figure \ref{fermi}.)
\begin{figure}[h]
    \centering
    \begin{tikzpicture}[scale=0.5]
    \draw (-5,0) node [below] {$a$};
    \draw (5,0) node [below] {$b$};
    \draw [dashed] (-5,0) -- (5,0);
    \draw [dashed] (-5,0) circle [radius=2];
    \draw [dashed] (5,0) circle [radius=2];
    \draw [blue] (4,1) arc [radius=1.5, start angle=135, end angle = 225];
    \draw [blue] (2,2) arc [radius=3, start angle=135, end angle = 225];
    \draw [blue] (0,2.5) arc [radius=3.5, start angle=135, end angle = 225];
    \draw [blue] (-2,3) arc [radius=4, start angle=135, end angle = 225];
    \draw (0,0) node [above] {$R$};
    \draw (5,2) node [above] {$\mathcal{R}_b$};
    \draw (-5,2) node [above] {$\mathcal{R}_a$};
    \draw [blue] (5,-5) node [below] {$t$, "Causal influence" of the signal};
    \draw [blue] [->] (5,-4) -- (0,-4);
    \end{tikzpicture}
    \caption{Fermi's two-atom system.}
    \label{fermi}
\end{figure}

Yngvason improved upon Fermi's result of the transition probability from $a$'s ground to $a$'s excited state using the observables available in QFT but not in QM. More specifically, since QFT does not have ``minimal projectors,'' we can't make a measurement that merely acts like a switch from ``unexcited'' to ``excited.''\footnote{Technical aside: Let $\A$ be a von Neumann algebra. Two orthogonal projectors $P_1, P_2 \in  \A$ are called equivalent with respect to $\A$, denoted $P_1 \sim P_2$, if there exists an operator $V\in\A$ such that $P_1 = V^*V$ and $P_2 = VV^*$. Projector equivalence is an equivalence relation on the projector lattice on the Hilbert space $ \Hcal$ on which the von Neumann algebra is represented. Denote $P_1\leq P_2$ when $P_1\sim P_1'$ and $P_1'\Hcal\subseteq P_2 \Hcal$. If $\A$ is a von Neumann factor, $\leq$ is a total order \citep[Theorem 2.1.9]{Haag1996}. A projector is said to be finite if it is not equivalent to any proper subprojector. The notion of finiteness, together with the total ordering, allows us to define a projector's relative dimension $d$, a positive number (possibly infinite) where $d(P)=0$ if and only if $P=0$, $d(P_1)\leq d(P_2)$ if and only if $P_1\leq P_2$, $d(P_1+P_2)=d(P_1)+d(P_2)$ if $P_1\perp P_2$ etc. (See \citep[$\S$ 7.6]{moretti_spectral_2017} for details). Then, a minimal projector $P$ has no non-zero subprojectors, so it is used to normalize $d(P)=1$. Equivalently, minimal projectors are the atoms of the projector lattice. That is the precise sense in which they are ``switch-like'' in footnote \ref{types}. Only type I factors contain minimal projectors.} However, we can still know of changes in the state of $a$ by purely local operations.\footnote{\label{types} Technical aside: Yngvason's ``strong local preparability'' only needs that the algebras are type III von Neumann factors, but we also want to prepare the states with arbitrary precision for any measurement to be as ``switch-like'' as possible. There are at least two strategies for doing this. One is Yngvason's, which uses the subclassification of type III von Neumann factors: type III$_1$ factors help specify further \textit{how} states can be prepared \citep[Theorem 4]{connes_homogeneity_1978}. Another one is to use the non-local type I algebra interpolating local type III algebras satisfying the split property to approximate local operations \citep[Corollary 1 of Theorem 4]{kitajima_local_2018}. Although the operational significance of this strategy has been questioned in the literature \citep[\S 5.2.5]{ruetsche_interpreting_2011}, it's not clear to me that it's totally unviable. Even if we need to be careful about using type I algebras to study relativistic causation \citep{Hegerfeldt1994}, results like Kitajima's or those in \citep{busch_unsharp_1999} show that rejecting them altogether would be too rash, and even Yngvason made a mistake in the argument ruling out using type I factors. Probably inspired by MC, he associates the orthogonality of $a$'s ground and excited states with the fact that their respective orthogonal projectors will localize them in causally independent regions. But, if $A$ is a self-adjoint operator of a von Neumann algebra $\A$ (independently of its type), all of $A$'s spectral projections belong to $\A$ \citep[Theorem 4.1.11]{murphy_calgebras_1990}, including orthogonal projections. Thanks to Aleksandr Pinzul for bringing this theorem to my attention.} The crux of the setup is to define two states, one in which atom $b$ is excited in $\mathcal{R}_b$ at $t=0$ and a state in which atom $b$ is in its ground state (or absent, since $a$ only ``knows" that some radiation came into $\mathcal{R}_a$, not the state of $b$) in $\mathcal{R}_b$ at $t=0$. Since these are joint states of the atoms, doing a measurement outside of $\mathcal{R}_b$ at $t=0$ cannot distinguish them from the ground state of $a$; one can only check whether $b$ is in its ground state or excited state at $t=0$ \textit{from inside $\mathcal{R}_b$}.

Then, to check whether $a$ became excited by the signal emitted by $b$ \textit{measured in $\mathcal{R}_a$} is to consider how much these states deviate from each other at some other time $t$ with observables in $\Ocal_a$ (where $\Ocal_a$ is the spacetime region made up of the spatial region $\mathcal{R}_a$ and an infinitesimal interval around $t=0$). Since $\Ocal_a''$ includes the region that the light rays have not yet reached, the two states are indistinguishable for $t<R/c$ if LPC holds. As soon as $b$ de-excites itself, it is irrelevant what happens to it from $a$'s point of view; it would not be able to detect any effect before the arrival of the light rays stemming from $b$'s de-excitation anyway. So LPC guarantees that $a$ doesn't know of $b$'s presence before it detects the radiation it emitted!

Notice that $\A(\Ocal_a)=\A(\Ocal_a'')$ implies that we have ``enlarged'' the region $\Ocal_a$ to $\Ocal_a''$. Since the light rays emitted from $b$ are moving, the region causally dependent on $\Ocal_b$ is growing. Only when the rays reach $\mathcal{R}_a$ will an observer be able to measure whether the state of $a$ changed. The light rays from PC's notion of causal dependence need not originate from points at rest at the reference frame at which their light cones are drawn, so here we see how a region flips in time from causally independent to causally dependent. So, aside from showing how LPC rules out superluminal signaling, Fermi's experiment helps to see how LPC finesses the notion of causal dependence, as argued in the previous subsection.

\subsection{Further advantages of LPC}

There's more. First, LPC can be generalized to curved spacetimes and used in concrete measurement protocols \citep{FV2020}, thus alleviating some of the worries I had raised about SC and MC.\footnote{However, I am not claiming that LPC solves problems like Sorkin's paradox since that would require a different conception of measurements in QFT from the one I have been assuming in this paper. The advantages I have presented from Buchholz, Yngvason, and Fewster and Verch rely on specific measurement protocols or experimental setups. Fewster and Verch do not assume LPC explicitly, but they obtain it for free (and with it, its advantages) because their framework relies heavily on \textit{causally convex} (i.e., diamond-like) regions, for which LPC is trivially true. Taking all regions to be diamonds is not esoteric; Haag uses it to formulate a lattice of causally complete regions that makes the connection between causal complements and algebraic commutants salient (\citeyear[\S III.4.1]{Haag1996}) and to study superselection sectors (\citeyear[p. 156]{Haag1996}). However, Fewster and Verch's version of LPC is more general than the one I presented here because they work with local $^*$-algebras, not von Neumann algebras, which gives them less topological structure than the one the derivation of LPC from the other axioms that I cite in the main text can exploit.} Second, in keeping this paper minimally technical, I have avoided any talk on the details of the mathematical structure of observable algebras because my arguments are independent of these considerations. However, these details are vital for relativistic causation. For example, superluminal signaling is unavoidable if the algebras are the matrix algebras of QM, which do have minimal projectors \citep{Hegerfeldt1994}. From this point of view, Buchholz and Yngvason (\citeyear{Buchholz1994, Yngvason2005}) have shown how LPC and the specific type of von Neumann algebra relevant for AQFT come together in implementing relativistic causation. (Cf. footnote \ref{types} for further discussion).

\subsection{\textsc{Holistic} redux: LPC is implied by the causal axioms}

The past subsections could have been the ingredients for a defense of \textsc{Atomistic}: LPC should supersede the three causal axioms as the most direct expression of relativistic causation in AQFT since it captures all the desiderata. However, this attempt at privileging one condition or even having it replace the rest would be too rash for two reasons. First, there are results important for the foundations of QFT, like the CPT theorem, that can be derived from the three causal axioms but that are logically independent of LPC \citep[\S 5.1]{Haag1996}. So even if LPC has some additional advantages compared to the three causal axioms \textit{from the point of view of the causal structure of the theory}, the axioms have more physical content than that which goes into codifying relativistic causation. Plus, I have been discussing in this paper which axioms do the causal heavy lifting for the theory, not which ones are necessary to derive the core theorems of QFT or to have a consistent theory.

However, more importantly for my purposes, we can derive LPC from the three causal axioms (and some other physically reasonable conditions), as I will sketch below as a defense of \textsc{Holistic}. Consider the construction of \citep{Haag1962} from figure \ref{diamond}.
\begin{figure}[h]
\centering
%0.45
\begin{tikzpicture}[scale=0.66]
\draw [->] (-7,0) -- (7,0);
\draw (7.5,0) node {$x$};
\draw [->] (0,-6) -- (0,6);
\draw (0,6.5) node {$t$};
\draw[pattern=south west lines] (-3,-2) -- (-3,2) -- (3,2) -- (3,-2) -- cycle;
\draw (-7,-2) -- (0,5);
\draw (-7,2) -- (0,-5);
\draw (7,-2) -- (0,5);
\draw (7,2) -- (0,-5);
\draw[very thick, dotted, red] (-7.5,2) -- (7.5,2);
\draw[very thick, dotted, red] (-7.5,-2) -- (7.5,-2);
\fill[pattern=vertical lines] (5,0) -- (7,2) -- (7,-2) -- (5,0);
\fill[pattern=vertical lines] (-5,0) -- (-7,2) -- (-7,-2) -- (-5,0);
\draw (0.5,4.8) node [right] {$\mathcal{D}$};
\draw (0,2.5) node [right] {$\mathcal{C}$};
\draw (3.6,0) node [above] {$\mathcal{C}_r$};
\draw (-3.6,0) node [above] {$\mathcal{C}_r$};
%\draw (0,3.5) node [left] {$\mathcal{C}_t$};
%\draw (0,-3.5) node [left] {$\mathcal{C}_t$};
\end{tikzpicture}
\caption{The region shaded with slanted lines is $\mathcal{C}$. The region shaded with vertical lines is $\mathcal{C}'$, the causal complement of $\mathcal{C}$. The diamond $\mathcal{D}$, containing $\mathcal{C}$, is the causal complement of $\mathcal{C}'$. The red dotted lines extend spatially to form a time slice of the temporal size of $\mathcal{C}$. Call $\mathcal{C}_r$ the two caps outside of $\mathcal{C}$ that extend in the spatial direction.}
\label{diamond}
\end{figure}%\clearpage
%\captionof{figure}{}
%\begin{figure}[h]
%\captionsetup{labelformat=empty}
%  \contcaption{}% Continued caption
%\end{figure}
It is now possible to prove that $\A(\mathcal{D})=\A(\mathcal{C})$, which is exactly LPC since the diamond $\mathcal{D}$ satisfies $\mathcal{D}=\mathcal{C}''$. All the pieces of the proof have been put together in the appendix, but here, I will only sketch the main steps. From SC, one can prove that the domain of analyticity of the fields in $\mathcal{C}$ can be extended to $\mathcal{C}\cup\mathcal{C}_r$ \citep{Borchers1961, Streater2000}. From PC and MC, we can build a time slice from $\mathcal{C}\cup\mathcal{C}_r$ and $\mathcal{C}'$, where $\mathcal{C}$ and $\mathcal{C}'$ localize separated subsystems. The fact that the observables at a time slice generate those in all spacetime gives some nice closure properties that display the intimate connection between the local algebras and the regions in which they are supported. Finally, from MC, the observables from $\mathcal{C}' $ commute with those of $\mathcal{C}$ but also with those of $\mathcal{C}'' (=\mathcal{D})$. The takeaway from this construction is that even if SC, MC, and PC individually failed to characterize relativistic causation in AQFT, LPC captures all the desiderata of relativistic causation in AQFT \textit{because} it is a fruit of their conjunction.

%\vspace{-1em}

\section{Outlook}

Throughout this paper, we saw that SC, MC, and PC only give partial characterizations of relativistic causation. Even if this is unsatisfactory for the advocates of \textsc{Atomistic}, the need for multiple axioms should not be startling anymore precisely because they highlight different aspects of the causal framework of the theory. Additionally, LPC encapsulates each axiom's advantages for the most widely used regions and theories in AQFT, which was the key idea behind my defense of \textsc{Holistic}. Assuming LPC has become a widespread move in more technical literature in AQFT. Still, its motivations are rarely stated, its role as a causal axiom is left uninterpreted, and its interdependence with the other causal axioms is ignored.

However, many additional worries emerged in analyzing this wide variety of axioms. First, an appropriate interpretation of MC would require tackling the measurement problem in QFT. One way to address this problem is to introduce a more sophisticated account of local operations in AQFT, which is where some of the more technical literature has been heading \citep{Okamura2016, kitajima_local_2018, Drago2020}. However, from an interpretational point of view, it is unclear whether considering a more widespread class of measurements would solve the problem of answering what ``measurement'' means. Even if looking for an account of measurements in QFT seems orthogonal to a project on relativistic causation, I do think they are connected through the worries that philosophers have raised about the ``operationalist'' views of the founders of AQFT and the ``algebraic imperialist'' \citep{Ruetsche2011} tendencies of the mathematical physicists working on this formulation of QFT. Additionally, revising the connections between causation and measurements would require further examining the local algebras' mathematical structure and the states that can act on them. The need to appeal to those tools can be seen not only in the problems of measurements in, e.g., MC but in the interplay between the different formulations of PC as a requirement demanding that we have a compatible notion of local dynamics and global determinism. That is one of the problems that even promoting LPC to an additional assumption in AQFT leaves unsettled.

The second problem is that any account of relativistic causation in AQFT is challenged by entanglement, which is ubiquitous and maximal in AQFT, even in spacelike separated regions \citep{Summers1988}. Entanglement \textit{itself} is not something we should worry about since it is just a key feature of QM and QFT. Instead, the origin of the uneasiness about entanglement is that it is often conceived as a form of non-locality or non-separability, which conflicts with, e.g., conceiving MC as a form of independence between spacelike separated subsystems. The pull of entanglement as a problem for causation in AQFT relies heavily on assumptions outside of the theory's axiomatic framework about measurements and operations in QFT \citep{Ruetsche2021} and, as such, it is connected to my first remaining worry. These problems are, to the best of my knowledge, unsettled. I hope this diagnostic points to some of the issues that need to be addressed and the importance of doing so.

\section*{Acknowledgements}

This paper and the many shapes it has taken have profited from helpful criticisms, comments, and conversations with many people, including various anonymous reviewers. Thanks to audiences at Universidad de los Andes, the University of Michigan, and the 2022 versions of the LMP Graduate Student Conference, BSPS, and PSA. Iván Burbano, Josh Hunt, Gabrielle Kerbel, and Yixuan Wu commented on earlier versions of the manuscript. Aleksandr Pinzul, Andrés Reyes, and Noel Swanson gave crucial technical guidance. Special thanks go to Dave Baker, Gordon Belot, Calum McNamara, and Laura Ruetsche for their diligent and sustained feedback and support.

\section*{Appendix: Causal diamond proof using essential duality}

\subsection*{Preliminaries: notation, some definitions, and re-stating the axioms}

Let $\Ocal$ be an open, bounded subregion of Minkowski spacetime $\mathcal{M}$, and $\A(\Ocal)$ a von Neumann algebra whose elements are taken to be the observables describing the physics in $\Ocal$. I'll take the algebras to be ``concretely represented,'' which means that $\A(\Ocal)\subseteq\B(\Hcal)$, the bounded, linear operators on some Hilbert space $\Hcal$. I'll also assume that $\A(\mathcal{M})=\B(\Hcal)$. Denote by $\A(\Ocal)'$ the \textbf{commutant} of $\A(\Ocal)$---the set of all operators commuting with those of $\A(\Ocal)$. A \textbf{von Neumann factor} is a von Neumann algebra satisfying $\A(\Ocal)\cup\A(\Ocal)'=\B(\Hcal)$ or, equivalently, $\A(\Ocal)'\cap\A(\Ocal)=\mathbb{C}\Id_{\B(\Hcal)}$.

Consider two spacetime regions $\Ocal_1$ and $\Ocal_2$. $\Ocal_2$ \textbf{depends causally} on $\Ocal_1$ if every light ray in the backward (or forward) light cone originating from any point in $\Ocal_2$ intersects $\Ocal_1$. If $\Ocal_1\subseteq\Ocal_2$, \textbf{isotony} tells us that $\A(\Ocal_1)\subseteq\A(\Ocal_2)$. Let $D(\Ocal)$ denote the \textbf{domain of dependence} of $\Ocal$, the locus of causally dependent points to $\Ocal$, and let $\Ocal'$, the \textbf{causal complement} of $\Ocal$, be the set of points outside $D(\Ocal)$. \textbf{Essential duality} \citep[p. 841]{Halvorson2007} then tells us that $\A(\mathcal{\Ocal}')'=\A(\mathcal{\Ocal}'')$.

The following \textbf{causal axioms} follow the notation and terminology from the main text.

\begin{itemize}
    \item \textbf{Spectrum Condition} (SC): (a) The joint spectrum of the infinitesimal generators of translations in Minkowski spacetime is confined to the forward lightcone. (b) There is a unique (up to phase factors) \textbf{vacuum state}, which is translationally invariant and has zero energy-momentum.
    \item \textbf{Microcausality} (MC): $\A(\Ocal_1)$ commutes with $\A(\Ocal_2)$ if $\Ocal_1$ and $\Ocal_2$ are space-like separated. In symbols, $\A(\Ocal)\subseteq\A(\Ocal')'$.
    \item \textbf{Primitive Causality} (PC): Consider a \textbf{time slice} in $\mathcal{M}$. That is, a region infinitely extended in space but restricted to a time interval of size $\tau$: $\Ocal_{\tau}:= \{x\in\mathcal{M}: \abs{x^0 - t} < \tau \}$. Then $\A(\Ocal_{\tau})=\A(\mathcal{M}) \ \forall\tau$.
\end{itemize}

\subsection*{Proof}

Consider the following cylinder $\mathcal{C}:=\{x\in\mathcal{M}:||\vec{x}||<a,|x^0|<\tau\}$ for some $a$ and $\tau$, shaded with diagonal lines in figure \ref{diamondp}. Its causal complement $\mathcal{C}'$ is shaded with vertical lines. Call the caps between the cylinder and the vertices of the diamond $\mathcal{C}_t$ and $\mathcal{C}_r$, according to their orientation toward the time or space axis. Let $\mathcal{D}$ be the diamond $\mathcal{C}\cup\mathcal{C}_t\cup\mathcal{C}_r$, which is equivalent to $\mathcal{C}''$. This proof aims to show that $\A(\mathcal{C})=\A(\mathcal{D})$. Although this result is only suited for regions such as the diamond, $\A(\Ocal)=\A(\Ocal'')$ is usually lifted to a general property. See the main text for further discussion.

\begin{figure}[h]
\centering
\begin{tikzpicture}[scale=0.66]
\draw [->] (-7,0) -- (7,0);
\draw (8,0) node {$||\vec{x}||$};
\draw [->] (0,-7) -- (0,7);
\draw (0,8) node {$x^0$};
\draw[pattern=south west lines] (-3,-2) -- (-3,2) -- (3,2) -- (3,-2) -- cycle;
\draw (-7,-2) -- (0,5);
\draw (-7,2) -- (0,-5);
\draw (7,-2) -- (0,5);
\draw (7,2) -- (0,-5);
\fill[pattern=vertical lines] (5,0) -- (7,2) -- (7,-2) -- (5,0);
\fill[pattern=vertical lines] (-5,0) -- (-7,2) -- (-7,-2) -- (-5,0);
\draw (0,3) node [right] {$\mathcal{C}_t$};
\draw (0,-3) node [right] {$\mathcal{C}_t$};
\draw (4,0) node [above] {$\mathcal{C}_r$};
\draw (-4,0) node [above] {$\mathcal{C}_r$};
\end{tikzpicture}
\caption{Causal diamond.}
\label{diamondp}
\end{figure}

Here's a sketch of the proof, where every step will have its own subproof.

\begin{center}
    \begin{enumerate}
        \item From SC, $\A(\mathcal{C}\cup\mathcal{C}_r)=\A(\mathcal{C})$.
        \item From PC and 1, $(\A(\mathcal{C}\cup\mathcal{C}_r)\cup\A(\mathcal{C}'))''=(\A(\mathcal{C})\cup\A(\mathcal{C}'))''=\B(\Hcal)$.
        \item From MC and 2, $\A(\mathcal{C})$ is a von Neumann factor.
        \item[$\therefore$] From MC and essential duality, $\A(\mathcal{C})=\A(\mathcal{D})$.
    \end{enumerate}
\end{center}

\subsubsection*{Proof of 1}

The proof of this step is the most elaborate one, and it relies crucially on a lemma found in \citep{Borchers1961}. However, Borchers doesn't clarify what the role of SC is, so, in \citep[pp. 22-31]{calderon_ossa_survey_2019}, where much of this appendix comes from, I supplemented Borchers' proof with results found in \citep{Streater2000}. I've also added more details to some of the proofs I adapted from them.

\begin{displayquote}
    \textbf{Streater and Wightman's Theorem 2-13:} Let $F_1$ be a function holomorphic in an open set $D_1$ in the upper half-plane with an open interval $a<x<b$ as part of its boundary. Let $F_2$ be holomorphic in an open set $D_2$ in the lower half-plane with $a<x<b$ as part of its boundary. Suppose $F_1(x)=\lim_{y\rightarrow0^+}F_1(x+iy)$ and $F_2(x)=\lim_{y\rightarrow0^+}F_2(x-iy)$ exist uniformly, in $a<x<b$, are continuous and satisfy $F_1(x)=F_2(x)$ for $a<x<b$. Then, $F_1$ and $F_2$ are holomorphic on $a<x<b$ and are the same holomorphic function. 
\end{displayquote}

\noindent The theorem, a special case of Bogolyubov's Edge of the Wedge theorem \citep{rudin_lectures_1971, Bogolubov1990, vladimirov_bogolyubovs_1994}, states the condition under which the analytic continuation of two holomorphic functions is such that they coincide in their real boundaries. We'll use this result below and, as Streater and Wightman discuss, this result can be extended for distributions whose real part is compactly supported.

Let $V_+$ be the set of all real momentum four-vectors $p$ such that $p^2>0$ and $p^0>0$ on some reference frame, and let $\bar{V}_+$ be its closure (the set of $p$s such that $p^2\geq 0$ and $p^0\geq 0$). Define $V_-$ and $\bar{V}_-$ analogously but for negative energies. 

\begin{displayquote}
    \textbf{Borchers' lemma:} Let $F^{\pm}(x+iy)$ be analytic functions with boundary values $F^{\pm}(x)=\lim_{\substack{y\to 0 \\ y \in \bar{V}_{\pm}}} F^{\pm}(x+iy)$ that exist as tempered distributions. Then, if $F^+(x)=F^-(x)$ for $x\in \mathcal{C}$, $F^+(x)=F^-(x)$ for $x\in \mathcal{C}\cup \mathcal{C}_r$.
\end{displayquote}

\noindent \textit{Sketch of the proof.} From the special case of the Edge of the Wedge theorem from above, $F^+(\zeta)$ and $F^-(\zeta)$ coincide in their domain of analyticity, which contains the cylinder $\mathcal{C}$. Then, a re-parametrization of $\zeta$ within allows us to extend the domain of analyticity of $F^{\pm}$, and we can prove that the new $\zeta$ lies in $\mathcal{C}\cup\mathcal{C}_r$.

Before proving Streater and Wightman's Theorem 3-2(b), the last ingredient we're missing to prove my premise 1, we need an intermediate result that uses SC.

\begin{displayquote}
    \textbf{Intermediate result \citep[pp. 91-92]{Streater2000}:} Let $U(a,\Lambda)$ be a unitary representation of the Poincaré group on a Hilbert space $\Hcal$, where $a$ is a four-vector and $\Lambda$ is a Lorentz transformation. Let $\Phi, \Psi\in\Hcal$ and $\bra{}\ket{}$ be $\Hcal$'s inner product. Then, $\int \dd a \ e^{-ip\cdot a} \bra{\Phi}U(a,\Id)\ket{\Psi}=0$ unless $p$ belongs to the spectrum of the energy-momentum operators.
\end{displayquote}

\begin{proof} Let us write $\bra{\Phi}\ket{\Psi}$ in terms of a basis\footnote{Streater and Wightman recognize that this derivation might not be rigorous (p. 91), but it can be made rigorous with different techniques (p. 92). I'll stick to this version since it's more intuitive and to me and the role of SC is clearer.} of momentum eigenstates $\ket{Q_\alpha}$: $\bra{\Phi}\ket{\Psi}=\sum_{\alpha} \int \dd Q \bra{\Phi}\ket{Q_\alpha}\bra{Q_\alpha}\ket{\Psi}$. Then, we have that
\begin{align*}
    \int \dd a \ e^{-ip\cdot a} \bra{\Phi}U(a,\Id)\ket{\Psi} &= \sum_{\alpha} \int \dd Q \int \dd a \ e^{-ip\cdot a} \bra{\Phi}\ket{Q_\alpha}\bra{Q_\alpha}U(a,\Id)\ket{\Psi}\\
    &= \sum_{\alpha} \int \dd Q \int \dd a \ e^{-i(p-Q)\cdot a} \bra{\Phi}\ket{Q_\alpha}\bra{Q_\alpha}\ket{\Psi}\\ &= (2\pi)^4 \sum_{\alpha} \int \dd Q \ \delta(Q-p) \bra{\Phi}\ket{Q_\alpha}\bra{Q_\alpha}\ket{\Psi}
\end{align*}

\noindent The second equality follows from the fact that $U(a,\Id)=e^{i\hat{P}^\mu a_\mu}$, where $\hat{P}$ is the energy-momentum operator, and that $\ket{Q_\alpha}$ is one of its eigenvectors. Then, since $\Phi$ and $\Psi$ are arbitrary states, $\int \dd a \ e^{-ip\cdot a} \bra{\Phi}U(a,\Id)\ket{\Psi}=0$ unless $p$ is equal to some $Q\in\bar{V}_+$.\end{proof}

\begin{displayquote}
    \textbf{Streater and Wightman's Theorem 3-2(b):} Let $\varphi_1,\varphi_2,...\varphi_n$ be any components of any fields and recall the definition of a \textbf{Wightman function}: $W^{(n)} (x_1, ..., x_n):= \bra{0} \varphi_1 (x_1) ... \varphi_n (x_n) \ket{0}$. Then there are tempered distributions $\mathcal{W}(\xi_1, ..., \xi_{n-1})$ depending on the coordinates $\xi_j = x_j - x_{j+1}$ for $j=1, 2, ..., n-1$ that satisfy
    \begin{equation*}
    W^{(n)}(x_1, ..., x_n)=\mathcal{W}(\xi_1, ..., \xi_{n-1}).
    \end{equation*}
    Their Fourier transforms are tempered distributions defined by
    \begin{align*}
    \Tilde{W}^{(n)}(p_1, ..., p_n) = \int \dd x_1 ... \dd x_n \ e^{i\sum_{j=1}^n p_j x_j} W^{(n)}(x_1, ..., x_n) \\ \Tilde{\mathcal{W}}(q_1, ..., q_{n-1}) = \int \dd\xi_1 ... \dd\xi_{n-1} \ e^{i\sum_{j=1}^{n-1} q_j \xi_j} \mathcal{W}(\xi_1, ..., \xi_{n-1})
    \end{align*}
    and are related by
    \begin{equation*}
    \Tilde{W}^{(n)}(p_1, ..., p_n) = (2\pi)^4 \delta\left(\sum_{j=1}^n p_j\right) \Tilde{\mathcal{W}}(p_1, p_1 + p_2, ..., p_1 + p_2 + ... + p_{n-1}).
    \end{equation*}
    Further, we have
    \begin{equation*}
    \Tilde{\mathcal{W}}(q_1, ..., q_{n-1}) = 0
    \end{equation*}
    if any $q$ is not in the energy-momentum spectrum of the states.
\end{displayquote}

\begin{proof}
The existence of $\mathcal{W}$ follows automatically from the fact that the Wightman functions are themselves tempered distributions \citep{wightman_how_1996} and that they're translationally invariant \citep[pp. 38-40]{Streater2000}. From their translational invariance, their Fourier transforms, straightforwardly generalized from functions to distributions, are well-defined.

As for the relationship between their Fourier transforms, it follows from the Fourier transform of the Dirac delta we already used in the intermediate result above and from the following trick: $\sum_{j=1}^n p_j x_j = p_1(x_1-x_2) + (p_1+p_2)(x_2-x_3) + ... + (p_1 + ... + p_{n-1})(x_{n-1} - x_n) + x_n\sum_{j=1}^n p_j = p_1\xi_1 + (p_1+p_2)\xi_2 + ... + (p_1 + ... + p_{n-1})\xi_{n-1} + x_n\sum_{j=1}^n p_j$.

As for the last result, let us use the intermediate result from above, with $\Phi\rightarrow\ket{0}$ and $U(a,\Id)\Psi\rightarrow\varphi_1 (x_1) ... \varphi_j (x_j) U(-a,\Id)\varphi_{j+1} (x_{j+1}) \varphi_n (x_n)\ket{0}=\varphi_1 (x_1) ... \varphi_j (x_j+a)\varphi_{j+1} (x_{j+1}) ... \varphi_n (x_n)\ket{0}$. That way, the inner product of $\Phi$ and $\Psi$ is directly related to the Wightman functions. The ``new'' $\Psi$ is also a unitary operator acting on a state, but we ``moved'' the operator using its unitarity and the translational invariance of the Wightman functions to obtain the following general result for $p\not\in\bar{V}_+$

\begin{equation*}
    \int\dd a \ e^{ip\cdot a} \mathcal{W}(\xi_1, ..., \xi_j+a, ..., \xi_{n-1}) = 0 \ , \ j = 1, 2, ..., n.
\end{equation*}

\noindent Since this result holds for any $j = 1, 2, ..., n$ and $a$ is an arbitrary translation, $\Tilde{\mathcal{W}}(q_1, ..., q_{n-1}) = 0$ unless \textit{each} $q_j$ lies in the physical spectrum. Using the relationship between the original Wightman functions and the translated ones, their Fourier transform vanishes unless $p_1, p_1+p_2,...,p_1+p_2+...+p_n\in\bar{V}_+$.
\end{proof}

We can now go back to Borchers' paper (\citeyear[pp. 791-792]{Borchers1961}) and prove my premise 1.

\textit{Rest of the proof.} Let $C$ be a bounded operator and let $P$ and $Q$ be eigenstates of the energy-momentum operator such that $[\varphi(x),C]=0$ for $\varphi(x)$ a scalar field and $x\in\mathcal{C}$. Consider the tempered distributions $F^+(x)=\bra{P}\varphi(x)C\ket{Q}$ and $F^-(x)=\bra{P}C\varphi(x)\ket{Q}$. (The distributions will only be tempered if $|P_0|,|Q_0|<\infty$, but this is expected from SC). From Streater and Wightman's theorem 3-2(b), their Fourier transforms satisfy $F^+(k) \neq 0$ only for $k+P \in \bar{V}_+$ and $F^-(k) \neq 0$ only for $k-Q \in \bar{V}_-$. Since $\varphi(x)$ and $C$ commute in $\mathcal{C}$, $F^+(x)=F^-(x)$ for $x\in \mathcal{C}$, but this implies $F^+(x)=F^-(x)$ for $x\in \mathcal{C}\cup\mathcal{C}_r$ due to Borchers' lemma from above. It follows that $[\varphi(x),C]=0$ for $x\in\mathcal{C}\cup\mathcal{C}_r$. Since the fields are the generators of the local algebras $\A(\Ocal)$ and since any bounded operator that commutes with fields in $\mathcal{C}$ also commutes with any field in $\mathcal{C}\cup\mathcal{C}_r$, $\A(\mathcal{C})'\subseteq\A(\mathcal{C}\cup\mathcal{C}_r)'$. Now, since $\A_1\subseteq\A_2$ implies $\A_2'\subseteq\A_1'$ for any two von Neumann algebras \citep[p. 114]{Haag1996}, we have that $\A(\mathcal{C})''\supseteq\A(\mathcal{C}\cup\mathcal{C}_r)''$. Since these are von Neumann algebras, $\A''=\A$, so $\A(\mathcal{C})\supseteq\A(\mathcal{C}\cup\mathcal{C}_r)$.\footnote{In the following pages, I will use the properties of von Neumann algebras from this sentence and the previous one without warning.} Due to the isotony property we have that $\A(\mathcal{C})\subseteq\A(\mathcal{C}\cup\mathcal{C}_r)$. Combining these results, we finally obtain $\A(\mathcal{C}\cup\mathcal{C}_r)=\A(\mathcal{C})$, which is precisely premise 1.

\subsubsection*{Proof of 2}

The proof of this step comes directly from \citep{Haag1962}. Let us take a time slice of the size of the cylinder, as shown in figure \ref{time-slice}.

\begin{figure}[h]
\centering
\begin{tikzpicture}[scale=0.5]
\draw [->] (-7,0) -- (7,0);
\draw (8,0) node {$||\vec{x}||$};
\draw [->] (0,-7) -- (0,7);
\draw (0,8) node {$x^0$};
\draw (-3,-2) -- (-3,2) -- (3,2) -- (3,-2) -- cycle;
\draw (-7,-2) -- (0,5);
\draw (-7,2) -- (0,-5);
\draw (7,-2) -- (0,5);
\draw (7,2) -- (0,-5);
\draw[very thick, dotted, red] (-8,2) -- (8,2);
\draw[very thick, dotted, red] (-8,-2) -- (8,-2);
\fill[pattern = south east lines] (-7,2) -- (-5,0) -- (-3,2);
\fill[pattern = south east lines] (7,2) -- (5,0) -- (3,2);
\fill[pattern = south east lines] (-7,-2) -- (-5,0) -- (-3,-2);
\fill[pattern = south east lines] (7,-2) -- (5,0) -- (3,-2);
\end{tikzpicture}
\caption{Time slice of the size of the cylinder delimited by the red dotted lines.}
\label{time-slice}
\end{figure}

From PC, the algebra of the time slice generates $\A(\mathcal{M})$, which contains all the bounded operators. Now, notice that the time slice is generated from $\mathcal{C}$, $\mathcal{C}_r$, a subset of $\mathcal{C}'$, and the caps between $\mathcal{C}'$ and $\mathcal{C} \cup \mathcal{C}_r$ shaded in the diagram above. Those caps, following \citep{Haag1962}, can be neglected if we take the smallest von Neumann algebra generated by $\A(\mathcal{C}\cup\mathcal{C}_r)$ and $\A(\mathcal{C}')$, $(\A(\mathcal{C}\cup\mathcal{C}_r)\cup\A(\mathcal{C}'))''$. We take the double commutant of the union (or, equivalently, their weak closure, following von Neumann's bicommutant theorem) since the union of two von Neumann algebras is not necessarily a von Neumann algebra. Since this is the smallest von Neumann algebra containing them, this choice should minimize the effect of ignoring the caps. Then, $(\A(\mathcal{C}\cup\mathcal{C}_r)\cup\A(\mathcal{C}'))''=\B(\Hcal)$, which is precisely premise 2. From premise 1, $\A(\mathcal{C}\cup\mathcal{C}_r)=\A(\mathcal{C})$, so we conclude that $(\A(\mathcal{C})\cup\A(\mathcal{C}'))''=\B(\Hcal)$.

\subsubsection*{Proof of 3}

The proof of this step follows automatically from MC and premise 2---so much so that Haag and Schroer don't include a proof! However, here is one. From MC, $\A(\mathcal{C}') \subseteq \A(\mathcal{C})'$. We can use this inclusion and premise 2 to prove that $\A(\mathcal{C})$ is a von Neumann factor.

\begin{proof}
\begin{align*}
\A(\mathcal{C}')& \subseteq \A(\mathcal{C})' \\ \A(\mathcal{C}) \cup \A(\mathcal{C}')& \subseteq \A(\mathcal{C}) \cup \A(\mathcal{C})' \\ (\A(\mathcal{C}) \cup \A(\mathcal{C}'))'& \supseteq (\A(\mathcal{C}) \cup \A(\mathcal{C})')' \\ (\A(\mathcal{C}) \cup \A(\mathcal{C}'))''& \subseteq (\A(\mathcal{C}) \cup \A(\mathcal{C})')'' \\ \B(\Hcal) = (\A(\mathcal{C}) \cup \A(\mathcal{C}'))''& \subseteq (\A(\mathcal{C}) \cup \A(\mathcal{C})')'' \subseteq \B(\Hcal) \\ (\A(\mathcal{C}) \cup \A(\mathcal{C})')''& = \B(\Hcal)
\end{align*}
\end{proof}

Since the set of von Neumann algebras on a Hilbert space forms an orthocomplemented lattice \citep[p. 114]{Haag1996}, taking the double commutant on the left-hand side of the last equation is the identity operation, so $\A(\mathcal{C}) \cup \A(\mathcal{C})'= \B(\Hcal)$, which is precisely premise 3.

\subsubsection*{Proof of the conclusion}

The proof of this final step was prepared as a supplement for the argument in the main text as an alternative ending from the version of the proof I had done in 2019 using Haag duality. I owe Noel Swanson the suggestion to try to use essential duality instead.

To prove that $\A(\mathcal{C})=\A(\mathcal{C}'')$, I will show the double inclusion.

\begin{itemize}
    \item One side follows from MC and essential duality: $\A(\mathcal{C})\subseteq\A(\mathcal{C}')'=\A(\mathcal{C}'')$.
    \item For the other side, our starting point is premise 3. Since $\A(\mathcal{C})$ is a von Neumann factor, $\A(\mathcal{C})\cup\A(\mathcal{C})'=\B(\Hcal)$ or, equivalently, that $\A(\mathcal{C})'\cap\A(\mathcal{C})=\mathbb{C}\Id_{\B(\Hcal)}$. Then, since $\A(\mathcal{C}')'\subseteq\B(\Hcal)=\A(\mathcal{C})\cup\A(\mathcal{C})'$, the elements of $\A(\mathcal{C}')'$ are in either $\A(\mathcal{C})$, which is what we want, or in $\A(\mathcal{C})'$. So we want to prove they are not in $\A(\mathcal{C})'$. Assume, for \textit{reductio}, that $\A(\mathcal{C}')'\subseteq\A(\mathcal{C})'$, which implies that $\A(\mathcal{C})''\subseteq\A(\mathcal{C}')''$. Since these are von Neumann algebras, $\A(\mathcal{C})''\subseteq\A(\mathcal{C}')''$ is equivalent to $\A(\mathcal{C})\subseteq\A(\mathcal{C}')$. Then, for some $A\in\A(\mathcal{C})$, $A\in\A(\mathcal{C}')$. Since $\A(\mathcal{C})$ is a von Neumann factor and $A\in\A(\mathcal{C})$, $A\not\in\A(\mathcal{C})'$ unless $A=c\Id_{\B(\Hcal)}$ for some $c\in\mathbb{C}$. For non-trivial operators, if $A\not\in\A(\mathcal{C})'$, $A\not\in\A(\mathcal{C}')''$ because $\A(\mathcal{C})'\subseteq\A(\mathcal{C}')''$ by MC. But, since these are von Neumann algebras, $\A(\mathcal{C'})''=\A(\mathcal{C'})$, so $A\not\in\A(\mathcal{C}')$, which is a contradiction. Therefore, $\A(\mathcal{C}')'\not\subseteq\A(\mathcal{C})'$. Since $\A(\mathcal{C}')'\subseteq\A(\mathcal{C})\cup\A(\mathcal{C})'$, we conclude that $\A(\mathcal{C}')'\subseteq\A(\mathcal{C})$. Using essential duality once again, $\A(\mathcal{C}'')\subseteq\A(\mathcal{C})$.
\end{itemize}

\bibliography{references}

\begin{thebibliography}{}

\bibitem[\protect\citeauthoryear{Bogolubov, Logunov, Oksak, and
  Todorov}{Bogolubov et~al.}{1987}]{Bogolubov1990}
Bogolubov, N.~N., A.~A. Logunov, A.~I. Oksak, and I.~Todorov (1987).
\newblock {\em General {Principles} of {Quantum} {Field} {Theory}}.
\newblock Mathematical {Physics} and {Applied} {Mathematics}. Springer
  Netherlands.

\bibitem[\protect\citeauthoryear{Bohr}{Bohr}{1948}]{Bohr1948}
Bohr, N. (1948).
\newblock {ON} {THE} {NOTIONS} {OF} {CAUSALITY} {AND} {COMPLEMENTARITY}.
\newblock {\em Dialectica\/}~{\em 2\/}(3/4), 312--319.

\bibitem[\protect\citeauthoryear{Borchers}{Borchers}{1961}]{Borchers1961}
Borchers, H.~J. (1961).
\newblock Über die {Vollständigkeit} lorentzinvarianter {Felder} in einer
  zeitartigen {Röhre}.
\newblock {\em Il Nuovo Cimento (1955-1965)\/}~{\em 19\/}(4), 787--793.

\bibitem[\protect\citeauthoryear{Bostelmann, Fewster, and Ruep}{Bostelmann
  et~al.}{2021}]{Bostelmann2021}
Bostelmann, H., C.~J. Fewster, and M.~H. Ruep (2021).
\newblock Impossible measurements require impossible apparatus.
\newblock {\em Physical Review D\/}~{\em 103\/}(2), 025017.

\bibitem[\protect\citeauthoryear{Buchholz and Yngvason}{Buchholz and
  Yngvason}{1994}]{Buchholz1994}
Buchholz, D. and J.~Yngvason (1994).
\newblock There are {No} {Causality} {Problems} for {Fermi}'s {Two} {Atom}
  {System}.
\newblock {\em Physical Review Letters\/}~{\em 73\/}(5), 613--616.

\bibitem[\protect\citeauthoryear{Busch}{Busch}{1999}]{busch_unsharp_1999}
Busch, P. (1999).
\newblock Unsharp localization and causality in relativistic quantum theory.
\newblock {\em Journal of Physics A: Mathematical and General\/}~{\em
  32\/}(37), 6535--6546.

\bibitem[\protect\citeauthoryear{Butterfield}{Butterfield}{2007}]{Butterfield2007}
Butterfield, J. (2007).
\newblock Reconsidering {Relativistic} {Causality}.
\newblock {\em International Studies in the Philosophy of Science\/}~{\em
  21\/}(3), 295--328.

\bibitem[\protect\citeauthoryear{Bär, Ginoux, and Pfäffle}{Bär
  et~al.}{2007}]{Bar2007}
Bär, C., N.~Ginoux, and F.~Pfäffle (2007).
\newblock {\em Wave {Equations} on {Lorentzian} {Manifolds} and
  {Quantization}}.
\newblock European Mathematical Society.

\bibitem[\protect\citeauthoryear{Calderón~Ossa}{Calderón~Ossa}{2019}]{calderon_ossa_survey_2019}
Calderón~Ossa, F. (2019).
\newblock A survey of causality in algebraic relativistic quantum field theory
  [{Undergraduate} thesis, {Universidad} de los {Andes}].

\bibitem[\protect\citeauthoryear{Connes and Størmer}{Connes and
  Størmer}{1978}]{connes_homogeneity_1978}
Connes, A. and E.~Størmer (1978).
\newblock Homogeneity of the state space of factors of type {III$_1$}.
\newblock {\em Journal of Functional Analysis\/}~{\em 28\/}(2), 187--196.

\bibitem[\protect\citeauthoryear{Curiel}{Curiel}{2017}]{curiel_primer_2017}
Curiel, E. (2017).
\newblock A {Primer} on {Energy} {Conditions}.
\newblock In D.~Lehmkuhl, G.~Schiemann, and E.~Scholz (Eds.), {\em Towards a
  {Theory} of {Spacetime} {Theories}}, pp.\  43--104. Springer.

\bibitem[\protect\citeauthoryear{Dimock}{Dimock}{1980}]{Dimock1980}
Dimock, J. (1980).
\newblock Algebras of local observables on a manifold.
\newblock {\em Communications in Mathematical Physics\/}~{\em 77\/}(3),
  219--228.

\bibitem[\protect\citeauthoryear{Dimock}{Dimock}{1982}]{Dimock1982}
Dimock, J. (1982).
\newblock Dirac quantum fields on a manifold.
\newblock {\em Transactions of the American Mathematical Society\/}~{\em
  269\/}(1), 133--147.

\bibitem[\protect\citeauthoryear{Drago and Moretti}{Drago and
  Moretti}{2020}]{Drago2020}
Drago, N. and V.~Moretti (2020).
\newblock The notion of observable and the moment problem for *-algebras and
  their {GNS} representations.
\newblock {\em Letters in Mathematical Physics\/}~{\em 110\/}(7), 1711--1758.

\bibitem[\protect\citeauthoryear{Duncan}{Duncan}{2012}]{duncan_dynamics_2012}
Duncan, A. (2012).
\newblock Dynamics {IV}: {Aspects} of locality: clustering, microcausality, and
  analyticity.
\newblock In {\em The {Conceptual} {Framework} of {Quantum} {Field} {Theory}},
  pp.\  132--170. Oxford University Press.

\bibitem[\protect\citeauthoryear{Earman}{Earman}{2014}]{Earman2014}
Earman, J. (2014).
\newblock No superluminal propagation for classical relativistic and
  relativistic quantum fields.
\newblock {\em Studies in History and Philosophy of Science Part B: Studies in
  History and Philosophy of Modern Physics\/}~{\em 48}, 102--108.

\bibitem[\protect\citeauthoryear{Earman and Valente}{Earman and
  Valente}{2014}]{EarmanValente2014}
Earman, J. and G.~Valente (2014).
\newblock Relativistic {Causality} in {Algebraic} {Quantum} {Field} {Theory}.
\newblock {\em International Studies in the Philosophy of Science\/}~{\em 28},
  1--48.

\bibitem[\protect\citeauthoryear{Einstein}{Einstein}{1948}]{Einstein1948}
Einstein, A. (1948).
\newblock {QUANTEN}-{MECHANIK} {UND} {WIRKLICHKEIT}.
\newblock {\em Dialectica\/}~{\em 2\/}(3/4), 320--324.

\bibitem[\protect\citeauthoryear{Epstein, Glaser, and Jaffe}{Epstein
  et~al.}{1965}]{Epstein1965}
Epstein, H., V.~Glaser, and A.~Jaffe (1965).
\newblock Nonpositivity of the energy density in quantized field theories.
\newblock {\em Il Nuovo Cimento (1955-1965)\/}~{\em 36\/}(3), 1016--1022.

\bibitem[\protect\citeauthoryear{Fermi}{Fermi}{1932}]{fermi_quantum_1932}
Fermi, E. (1932).
\newblock Quantum {Theory} of {Radiation}.
\newblock {\em Reviews of Modern Physics\/}~{\em 4\/}(1), 87--132.

\bibitem[\protect\citeauthoryear{Fewster}{Fewster}{2016}]{Fewster2016}
Fewster, C.~J. (2016).
\newblock The split property for quantum field theories in flat and curved
  spacetimes.
\newblock {\em Abhandlungen aus dem Mathematischen Seminar der Universität
  Hamburg\/}~{\em 86\/}(2), 153--175.

\bibitem[\protect\citeauthoryear{Fewster and Verch}{Fewster and
  Verch}{2020}]{FV2020}
Fewster, C.~J. and R.~Verch (2020).
\newblock Quantum {Fields} and {Local} {Measurements}.
\newblock {\em Communications in Mathematical Physics\/}~{\em 378\/}(2),
  851--889.

\bibitem[\protect\citeauthoryear{Fraser}{Fraser}{2011}]{fraser_how_2011}
Fraser, D. (2011).
\newblock How to take particle physics seriously: {A} further defence of
  axiomatic quantum field theory.
\newblock {\em Studies in History and Philosophy of Science Part B: Studies in
  History and Philosophy of Modern Physics\/}~{\em 42\/}(2), 126--135.

\bibitem[\protect\citeauthoryear{Geroch}{Geroch}{2011}]{Geroch2011}
Geroch, R. (2011).
\newblock Faster {Than} {Light}?
\newblock In {\em Advances in {Lorentzian} {Geometry}: {Proceedings} of the
  {Lorentzian} {Geometry} {Conference} in {Berlin}}, Volume~49 of {\em
  {AMS}/{IP} {Studies} in {Advanced} {Mathematics}}, pp.\  59--69. American
  Mathematical Society.

\bibitem[\protect\citeauthoryear{Haag}{Haag}{2010}]{haag_discussion_2010}
Haag, R. (1957/2010).
\newblock Discussion of the ‘axioms’ and the asymptotic properties of a
  local field theory with composite particles.
\newblock {\em The European Physical Journal H\/}~{\em 35\/}(3), 243--253.

\bibitem[\protect\citeauthoryear{Haag}{Haag}{1996}]{Haag1996}
Haag, R. (1996).
\newblock {\em Local {Quantum} {Physics}: {Fields}, {Particles}, {Algebras}\/}
  (2nd ed.).
\newblock Springer.

\bibitem[\protect\citeauthoryear{Haag and Schroer}{Haag and
  Schroer}{1962}]{Haag1962}
Haag, R. and B.~Schroer (1962).
\newblock Postulates of {Quantum} {Field} {Theory}.
\newblock {\em Journal of Mathematical Physics\/}~{\em 3\/}(2), 248--256.

\bibitem[\protect\citeauthoryear{Halvorson}{Halvorson}{2001}]{Halvorson2001}
Halvorson, H. (2001).
\newblock Reeh-{Schlieder} {Defeats} {Newton}-{Wigner}: {On} {Alternative}
  {Localization} {Schemes} in {Relativistic} {Quantum} {Field} {Theory}.
\newblock {\em Philosophy of Science\/}~{\em 68\/}(1), 111--133.

\bibitem[\protect\citeauthoryear{Halvorson}{Halvorson}{2007}]{Halvorson2007}
Halvorson, H. (2007).
\newblock Algebraic {Quantum} {Field} {Theory} (with an {Appendix} by {Michael}
  {Müger}).
\newblock In J.~Butterfield and J.~Earman (Eds.), {\em Philosophy of
  {Physics}}, pp.\  731--864. North-Holland.

\bibitem[\protect\citeauthoryear{Hegerfeldt}{Hegerfeldt}{1994}]{Hegerfeldt1994}
Hegerfeldt, G.~C. (1994).
\newblock Causality problems for {Fermi}’s two-atom system.
\newblock {\em Physical Review Letters\/}~{\em 72\/}(5), 596--599.

\bibitem[\protect\citeauthoryear{Hofer-Szabó and Vecsernyés}{Hofer-Szabó and
  Vecsernyés}{2018}]{HSV2018}
Hofer-Szabó, G. and P.~Vecsernyés (2018).
\newblock Locality and {Causality} {Principles}.
\newblock In {\em Quantum {Theory} and {Local} {Causality}}, pp.\  17--23.
  Springer.

\bibitem[\protect\citeauthoryear{Horuzhy}{Horuzhy}{1990}]{horuzhy_introduction_1990}
Horuzhy, S.~S. (1990).
\newblock {\em Introduction to {Algebraic} {Quantum} {Field} {Theory}}.
\newblock Kluwer Academic Publishers.

\bibitem[\protect\citeauthoryear{Kitajima}{Kitajima}{2018}]{kitajima_local_2018}
Kitajima, Y. (2018).
\newblock Local {Operations} and {Completely} {Positive} {Maps} in {Algebraic}
  {Quantum} {Field} {Theory}.
\newblock In M.~Ozawa, J.~Butterfield, H.~Halvorson, M.~Rédei, Y.~Kitajima,
  and F.~Buscemi (Eds.), {\em Reality and {Measurement} in {Algebraic}
  {Quantum} {Theory}}, pp.\  83--95. Springer Singapore.

\bibitem[\protect\citeauthoryear{Moretti}{Moretti}{2017}]{moretti_spectral_2017}
Moretti, V. (2017).
\newblock {\em Spectral {Theory} and {Quantum} {Mechanics} : {Mathematical}
  {Foundations} of {Quantum} {Theories}, {Symmetries} and {Introduction} to the
  {Algebraic} {Formulation}\/} (2nd ed.).
\newblock Springer.

\bibitem[\protect\citeauthoryear{Murphy}{Murphy}{1990}]{murphy_calgebras_1990}
Murphy, G.~J. (1990).
\newblock {\em C*–{Algebras} and {Operator} {Theory}}.
\newblock San Diego: Academic Press.

\bibitem[\protect\citeauthoryear{Okamura and Ozawa}{Okamura and
  Ozawa}{2016}]{Okamura2016}
Okamura, K. and M.~Ozawa (2016).
\newblock Measurement theory in local quantum physics.
\newblock {\em Journal of Mathematical Physics\/}~{\em 57\/}(1), 015209.

\bibitem[\protect\citeauthoryear{Papageorgiou and Fraser}{Papageorgiou and
  Fraser}{2023}]{papageorgiou_eliminating_2023}
Papageorgiou, M. and D.~Fraser (2023).
\newblock Eliminating the "impossible": {Recent} progress on local measurement
  theory for quantum field theory.
\newblock arXiv:2307.08524.

\bibitem[\protect\citeauthoryear{Peskin and Schroeder}{Peskin and
  Schroeder}{1995}]{Peskin1995}
Peskin, M.~E. and D.~V. Schroeder (1995).
\newblock {\em An {Introduction} to {Quantum} {Field} {Theory}}.
\newblock Westview Press.

\bibitem[\protect\citeauthoryear{Rudin}{Rudin}{1971}]{rudin_lectures_1971}
Rudin, W. (1971).
\newblock {\em Lectures on the {Edge}-of-the-{Wedge} {Theorem}}, Volume~6 of
  {\em {CBMS} {Regional} {Conference} {Series} in {Mathematics}}.
\newblock Providence, Rhode Island: American Mathematical Society.

\bibitem[\protect\citeauthoryear{Ruetsche}{Ruetsche}{2011}]{Ruetsche2011}
Ruetsche, L. (2011).
\newblock {\em Interpreting {Quantum} {Theories}}.
\newblock Oxford University Press.

\bibitem[\protect\citeauthoryear{Ruetsche}{Ruetsche}{2021}]{Ruetsche2021}
Ruetsche, L. (2021).
\newblock Locality in ({Axiomatic}) {Quantum} {Field} {Theory}: {A} {Minority}
  {Report}.
\newblock In E.~Knox and A.~Wilson (Eds.), {\em The {Routledge} {Companion} to
  {Philosophy} of {Physics}}, pp.\  311--322. Routledge.

\bibitem[\protect\citeauthoryear{Ruetsche and Earman}{Ruetsche and
  Earman}{2011}]{ruetsche_interpreting_2011}
Ruetsche, L. and J.~Earman (2011).
\newblock Interpreting {Probabilities} in {Quantum} {Field} {Theory} and
  {Quantum} {Statistical} {Mechanics}.
\newblock In C.~Beisbart and S.~Hartmann (Eds.), {\em Probabilities in
  {Physics}}, pp.\  263--290. Oxford University Press.

\bibitem[\protect\citeauthoryear{Rédei}{Rédei}{2014}]{redei_categorial_2014}
Rédei, M. (2014).
\newblock A categorial approach to relativistic locality.
\newblock {\em Studies in History and Philosophy of Science Part B: Studies in
  History and Philosophy of Modern Physics\/}~{\em 48}, 137--146.

\bibitem[\protect\citeauthoryear{Rédei and Valente}{Rédei and
  Valente}{2010}]{RedeiValente2010}
Rédei, M. and G.~Valente (2010).
\newblock How local are local operations in local quantum field theory?
\newblock {\em Studies in History and Philosophy of Science Part B: Studies in
  History and Philosophy of Modern Physics\/}~{\em 41\/}(4), 346--353.

\bibitem[\protect\citeauthoryear{Sorkin}{Sorkin}{1993}]{Sorkin1993}
Sorkin, R. (1993).
\newblock Impossible {Measurements} on {Quantum} {Fields}.
\newblock In B.~Hu and T.~Jacobson (Eds.), {\em “{Directions} in {General}
  {Relativity}”, {Proceedings} of the 1993 {International} {Symposium},
  {Maryland}, {Vol}. {II}: {Papers} in honor of {Dieter} {Brill}}, pp.\
  293--305. Cambridge UP.

\bibitem[\protect\citeauthoryear{Streater and Wightman}{Streater and
  Wightman}{2000}]{Streater2000}
Streater, R.~F. and A.~S. Wightman (1964/2000).
\newblock {\em {PCT}, {Spin} and {Statistics}, and {All} {That}}.
\newblock Princeton University Press.

\bibitem[\protect\citeauthoryear{Summers and Werner}{Summers and
  Werner}{1988}]{Summers1988}
Summers, S.~J. and R.~Werner (1988).
\newblock Maximal violation of {Bell's} inequalities for algebras of
  observables in tangent spacetime regions.
\newblock {\em Annales de l'I.H.P. Physique th\'eorique\/}~{\em 49\/}(2),
  215--243.

\bibitem[\protect\citeauthoryear{Swanson}{Swanson}{2017}]{swanson_philosophers_2017}
Swanson, N. (2017).
\newblock A philosopher's guide to the foundations of quantum field theory.
\newblock {\em Philosophy Compass\/}~{\em 12\/}(5), e12414.

\bibitem[\protect\citeauthoryear{Valente}{Valente}{2014}]{valente_does_2014}
Valente, G. (2014).
\newblock Does the {Reeh}–{Schlieder} theorem violate relativistic causality?
\newblock {\em Studies in History and Philosophy of Science Part B: Studies in
  History and Philosophy of Modern Physics\/}~{\em 48}, 147--155.

\bibitem[\protect\citeauthoryear{Vladimirov, Zharinov, and Sergeev}{Vladimirov
  et~al.}{1994}]{vladimirov_bogolyubovs_1994}
Vladimirov, V.~S., V.~V. Zharinov, and A.~G. Sergeev (1994).
\newblock Bogolyubov's “edge of the wedge” theorem, its development and
  applications.
\newblock {\em Russian Mathematical Surveys\/}~{\em 49\/}(5), 51.

\bibitem[\protect\citeauthoryear{Wallace}{Wallace}{2006}]{wallace_defence_2006}
Wallace, D. (2006).
\newblock In {Defence} of {Naiveté}: {The} {Conceptual} {Status} of
  {Lagrangian} {Quantum} {Field} {Theory}.
\newblock {\em Synthese\/}~{\em 151\/}(1), 33--80.

\bibitem[\protect\citeauthoryear{Wallace}{Wallace}{2011}]{Wallace2011}
Wallace, D. (2011).
\newblock Taking particle physics seriously: {A} critique of the algebraic
  approach to quantum field theory.
\newblock {\em Studies in History and Philosophy of Science Part B: Studies in
  History and Philosophy of Modern Physics\/}~{\em 42\/}(2), 116--125.

\bibitem[\protect\citeauthoryear{Weinberg}{Weinberg}{1995}]{Weinberg1995}
Weinberg, S. (1995).
\newblock Quantum {Fields} and {Antiparticles}.
\newblock In {\em The {Quantum} {Theory} of {Fields}}, Volume~1, pp.\
  191--258. Cambridge University Press.

\bibitem[\protect\citeauthoryear{Wightman}{Wightman}{1996}]{wightman_how_1996}
Wightman, A.~S. (1996).
\newblock How {It} {Was} {Learned} that {Quantized} {Fields} {Are}
  {Operator}-{Valued} {Distributions}.
\newblock {\em Fortschritte der Physik/Progress of Physics\/}~{\em 44\/}(2),
  143--178.

\bibitem[\protect\citeauthoryear{Yngvason}{Yngvason}{2005}]{Yngvason2005}
Yngvason, J. (2005).
\newblock The {Role} of {Type} {III} {Factors} in {Quantum} {Field} {Theory}.
\newblock {\em Reports on Mathematical Physics\/}~{\em 55\/}(1), 135--147.

\end{thebibliography}

\end{document}